\documentclass[10pt,final,doublecolumn]{IEEEtran}
\usepackage{amsmath}
\usepackage{latexsym}
\usepackage{graphicx}
\usepackage{bbding}
\usepackage{indentfirst}
\usepackage{algorithm,algorithmic}
\usepackage{setspace}
\usepackage{float}
\usepackage{epstopdf}
\usepackage{amssymb}
\usepackage{amsfonts}
\usepackage{enumerate}
\usepackage{multicol}
\usepackage{color}
\usepackage{slashbox}
\usepackage{bm}
\usepackage{amssymb}
\usepackage{stfloats}
\usepackage{epstopdf}
\usepackage{threeparttable}
\usepackage{epstopdf}
\usepackage{threeparttable}
\usepackage{subfigure}
\usepackage{makecell}

\usepackage{cite}

\IEEEoverridecommandlockouts
\allowdisplaybreaks[4]

\begin{document}
\title{6D Movable Antenna Based on User Distribution: Modeling and Optimization}

\author{{Xiaodan Shao,~\IEEEmembership{Member,~IEEE}, Qijun Jiang, and Rui Zhang, \IEEEmembership{Fellow, IEEE}}

	 \thanks{X. Shao is with the Institute for Digital Communications, Friedrich-Alexander-University Erlangen-Nuremberg, 91054
Erlangen, Germany, (e-mail: xiaodan.shao@fau.de).}

	 \thanks{Q. Jiang is with the School of Science and Engineering, Chinese University of Hong Kong, Shenzhen, China 518172, (e-mail: qijunjiang@link.cuhk.edu.cn).}

\thanks{R. Zhang is with School of Science and Engineering, Shenzhen Research Institute of Big Data, The Chinese University of Hong Kong, Shenzhen, Guangdong 518172, China. He is also with the Department of Electrical and Computer Engineering, National University of Singapore, Singapore 117583 (e-mail: elezhang@nus.edu.sg).
}
}
\maketitle

\IEEEpeerreviewmaketitle

\begin{abstract}
In this paper, we propose a new six-dimensional (6D) movable antenna (6DMA) system for future wireless networks to improve the communication performance. Unlike the traditional fixed-position antenna (FPA) and existing fluid antenna/two-dimensional (2D) movable antenna (FA/2DMA) systems that adjust the
positions of antennas only, the proposed 6DMA system consists of distributed antenna surfaces with independently adjustable three-dimensional (3D) positions as well as 3D rotations within a given space. In particular, this paper applies the 6DMA to the base station (BS) in wireless networks to provide full
degrees of freedom (DoFs) for the BS to adapt to the dynamic user spatial distribution in the network. However, a challenging new problem arises on how to optimally control the 6D positions and rotations of all 6DMA surfaces at the BS to maximize the network capacity based on the user spatial distribution, subject to the practical constraints on 6D antennas' movement.
To tackle this problem, we first model the 6DMA-enabled BS and the user channels with the BS in terms of 6D positions and rotations of all 6DMA surfaces. Next, we propose an efficient alternating optimization algorithm to search for the best 6D positions and rotations of all 6DMA surfaces by leveraging the Monte Carlo simulation technique. Specifically, we sequentially optimize the 3D position/3D rotation of each 6DMA surface with those of the other surfaces fixed in an iterative manner.
Numerical results show that our proposed 6DMA-BS can significantly improve the network capacity as compared to the benchmark BS architectures
with FPAs or MAs with limited/partial movability, especially when the user distribution is more spatially non-uniform.
\end{abstract}


\section{Introduction}
Mobile communications have come a long way and there was never an absence of innovative technologies when a new generation of wireless networks was introduced to succeed the preceding one. Over the last several decades, multi-antenna or so-called multiple-input multiple-output (MIMO) communication technologies have been significantly advanced to enable various generations of wireless networks.
By employing multiple antennas at the base station (BS) as well as user terminals, MIMO systems have provided substantial spatial
multiplexing and diversity gains to dramatically enhance the transmission rate and reliability of wireless systems \cite{ruimimo, mimo3}. In particular, MIMO technologies have evolved from single-user MIMO to multiuser MIMO, and subsequently massive MIMO at present with increasingly more antennas employed at the BS, which, however, results in its higher hardware cost and power consumption \cite{SP8,LA10}. To alleviate this issue, continuous-aperture MIMO, lens-antenna MIMO, and holographic MIMO technologies have been proposed to use subwavelength metamaterials to achieve superior beamforming performance cost-effectively \cite{con9, zenglens, holo}. Alternatively, there has also been a growing interest recently in leveraging intelligent reflecting surface (IRS) with passive reflection/beamforming \cite{qings,proc,shaos,nsr,shaotarget} to enhance the MIMO communication/sensing performance without mounting more active antennas at the BS. On the other hand, to improve the wireless network coverage, cooperative MIMO techniques by exploiting the joint signal transmission/reception among neighboring BSs, such as networked MIMO \cite{net},
coordinated multi-point (CoMP) \cite{comp}, and more recently cell-free massive MIMO \cite{free} have been introduced and thoroughly investigated.
Furthermore, as wireless systems have been migrating into higher frequency bands, millimeter wave (mmWave) MIMO has been proposed to provide high beamforming gains to compensate for the severe mmWave path loss \cite{m1}, leading to a renewed interest in analog beamforming and the hybrid analog/digital beamforming to balance between the system cost/complexity and beamforming performance \cite{2m,5m}.
\begin{figure}[t!]
\centering
\setlength{\abovecaptionskip}{0.cm}
\includegraphics[width=3.56in]{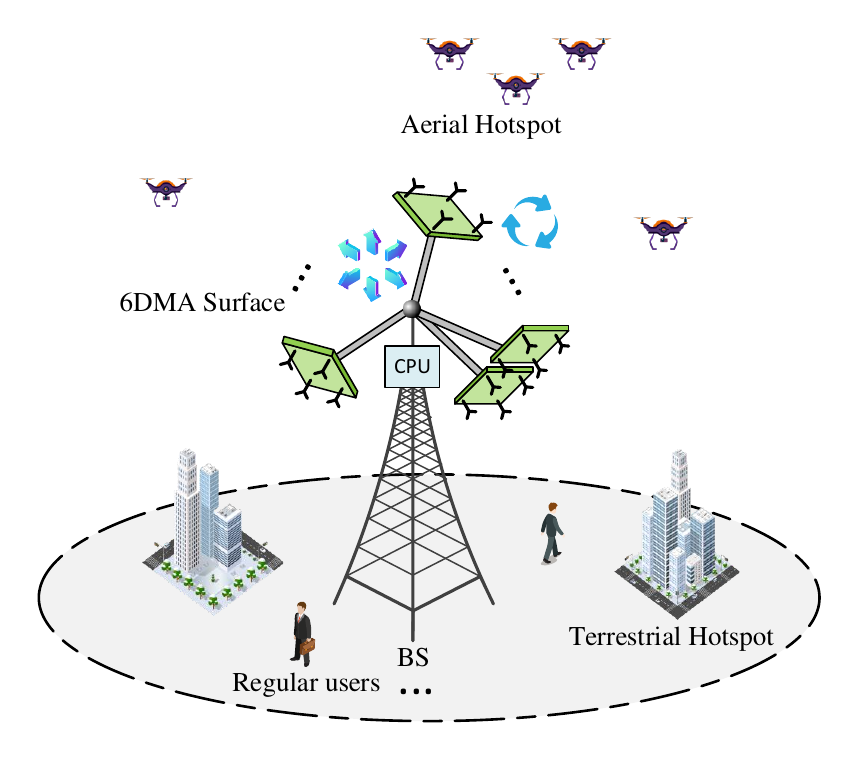}
\caption{6DMA-enabled BS for adapting to non-uniform user distribution in wireless networks.}
\label{practical_scenario}
\end{figure}

Although the above MIMO technologies have been extensively studied in the literature, they are all based on the traditional fixed-position antennas (FPAs) that cannot change their positions once deployed.
In contrast, the locations of mobile users in wireless networks change over time, thus leading to time-varying channels with their serving BS. To cope with such dynamic user channels, traditional wireless systems have adopted adaptive MIMO beamforming/processing as well as other transmission and resource allocation techniques \cite{gold,david}. However, due to the use of FPAs, their spatial degrees of freedom (DoFs) are fixed, and as a result, the BS cannot adapt to the dynamic user spatial distribution in the network efficiently. For example, as shown in Fig. \ref{practical_scenario}, a BS serves users in a given cell consisting of both terrestrial and aerial hot-spot areas with much higher user/device densities than the rest of the cell. In such scenarios,  the BS with FPAs (e.g., sector antennas) cannot flexibly allocate its spatial DoFs to match the non-uniform user distribution for achieving the maximum network capacity. To tackle this issue, existing BSs have been endowed with certain flexibilities in adjusting their antenna azimuth/tilt angles to adapt to the user distribution \cite{til3, tilt}.
However, such adjustments can only be applied to the whole antenna array (e.g., sector antenna) and are limited to either horizontal or vertical rotation, which thus cannot exploit the full spatial DoFs of all antennas at the BS.
\begin{table*}[!t]
\small
	\caption{{Comparison of different antenna architectures.}}
	\label{Table1}
	\centering
	\begin{tabular}{|c|c|c|c|c|c|c|c|}
		\hline
		\makecell{Antenna Architecture} & \makecell{Movability} & \makecell{Movable Unit} &\makecell{Movement Speed/Frequency} &\makecell{Performance \\Gain}
\\
		\hline
		6DMA & \makecell{High (position and rotation)}   & \makecell{Single antenna\\
/antenna surface }  & \makecell{Low \\(based on user distribution/statistical channel)}   & Very high \\
		\hline
		\makecell{FAS \cite{9264694,9650760}\\/2DMA \cite{wu,10243545,yifei}}  &Medium (position only)  & Single antenna  &\makecell{High \\(based on small-scale/instantaneous channel) }& High\\
		\hline
		\makecell{FPA \cite{til3, tilt}} & Low (azimuth/tilt angle only)  & Sector antenna  &\makecell{Low \\(based on user distribution)}& Low\\
\hline
	\end{tabular}
\end{table*}

To enable the BS's full flexibility in antenna deployment, we propose in this paper a new 6D movable antenna (6DMA) architecture as shown in Fig. \ref{practical_scenario}, where a set of 6DMA surfaces\footnote{For simplicity, we assume that each 6DMA surface is a uniform planar array (UPA) with a given size, while it can also take other shapes, such as curved/conformal surfaces.} are equipped at the BS, which can be independently adjusted in terms of both 3D positions and 3D rotations. Specifically, each
6DMA surface is connected with the central processing unit (CPU) at the BS via a separate rod, which is extendable and rotatable, and contains flexible wires (e.g., coaxial cable) that provide power supply to the 6DMA surface as well as enable the control/radio frequency (RF) signal exchange between it and the CPU. In addition, two motors are mounted at the two ends of the rod and controlled by the CPU to adjust the position and rotation of each 6DMA surface, respectively. Thereby, the BS can jointly design and control the 3D positions and 3D rotations of all 6DMA surfaces according to the spatial user distribution to maximize the multiuser-MIMO capacity (see Fig. \ref{practical_scenario}).

It is worth noting that the 6DMA system proposed in this paper differs significantly from the existing fluid antenna system (FAS) \cite{9650760,9264694} as well as 2D movable antenna (2DMA) system \cite{wu,10243545,yifei,qingmove}.
Firstly, FAS/2DMA can only adjust the positions of antennas within a given line/surface. In contrast, the proposed 6DMA system can adjust both the
3D position and 3D rotation of each antenna surface in the 3D space. With such 6D movability, the proposed 6DMA system can position and rotate antenna surfaces more flexibly to match the user distribution as shown in Fig. \ref{practical_scenario} for improving the network capacity.
Secondly, single antenna is adopted as the movable unit in the existing works on FAS/2DMA system (e.g., \cite{9650760,9264694,qingmove,wu,10243545,yifei}) to maximally exploit the small-scale channel spatial variation in the given region. This thus requires high-speed and frequent movement of the antennas in fast fading channels, which incurs high implementation cost and complexity\cite{review}.
In contrast, each 6DMA surface moves much more slowly and much less frequently for adapting to the large-scale fading channels of users, which change only when the spatial user distribution or statistical channel state information (CSI) in the network varies significantly. A summary of the above comparison of the proposed 6DMA and existing FAS/2DMA as well as traditional FPA is given in Table I.

The main contributions of this paper are summarized as follows.
\begin{itemize}
\item
Since directional antennas are used at BSs in current wireless networks \cite{3gpp}, their channels with the mobile users depend on not only the users' locations and their residing propagation environments, but also the antennas' 3D positions and 3D rotations at the BS. To characterize this effect, we first present a new 6DMA-enabled BS architecture with 6DMA surfaces each composed of an array of directional antenna elements and then model their channels with the user at a given location in terms of 3D position and 3D rotation of each 6DMA surface. Moreover, we introduce a general model of spatial user distribution in the network based on the non-homogeneous Poisson point process (NHPP) for evaluating the performance of the proposed 6DMA system.

\item
Next, we formulate an optimization problem to maximize the average network capacity with the 6DMA-BS by jointly designing the 3D positions and 3D rotations of all 6DMA surfaces, based on the knowledge of user spatial distribution. In particular, we consider practical deployment/movement constraints of 6DMAs including the minimum-distance constraint between any two 6DMA surfaces, as well as their rotation constraints for avoiding mutual signal reflection and signal blockage by the CPU of the BS. The intricate couplings among position and rotation variables of 6DMA surfaces, together with the aforementioned non-convex distance/rotation constraints, make this optimization problem non-convex and challenging to be solved optimally. To tackle this problem, we first apply the Monte Carlo simulation technique to approximate the average network capacity with a finite number of channel samples, and then propose an alternating optimization algorithm to maximize the approximate network
capacity efficiently. Specifically, the algorithm sequentially optimizes the 3D position/3D rotation of each 6DMA
surface with those of the other surfaces fixed in an iterative manner until the convergence is reached.

\item
Finally, we evaluate the performance of our proposed
6DMA-BS design and joint position/rotation optimization algorithm via numerical results. The
results demonstrate that the proposed design and algorithm can significantly improve the network capacity over
the benchmark BS architectures with the existing FPAs or MAs with limited or partial movability.
In particular, it is shown that the network capacity gain by 6DMA-BS becomes more appealing when the user spatial distribution exhibits more non-uniform and clustering (hot-spot) patterns.
\end{itemize}

The rest of this paper is organized as follows. Section II presents the
6DMA-BS architecture, the corresponding channel model, and the user spatial distribution. Section III formulates the network capacity maximization problem by jointly designing all 6DMA surfaces' positions and rotations under practical constraints.
Section IV presents the algorithm for solving the formulated problem by leveraging the Monte Carlo simulation and alternating optimization techniques. Section V presents numerical results for performance evaluation and comparison. Finally, Section VI concludes this paper.

\emph{Notations}: Boldface upper-case and lower-case letters denote
matrices and vectors, respectively, $(\cdot)^*$, $(\cdot)^H$, and $(\cdot)^T$  respectively denote conjugate, conjugate transpose, and transpose, $\mathbb{E}[\cdot]$ denotes the expected value of
random variable,  $\left \|\cdot\right \|_2$ denotes the Euclidean norm, $\mathbf{0}_{N}$ denotes the $N\times 1$ vector with all zero elements, $\mathbf{I}_N$ denotes the $N\times N$ identity matrix, $\mathrm{diag}({\bf x})$ denotes a diagonal matrix with the diagonal entries specified by vector ${\bf x}$, $[\mathbf{a}]_j$ denotes the $j$-th element of vector $\mathbf{a}$, $[\mathbf{A}]_{i,j}$ denotes the element of matrix $\mathbf{A}$ at the $i$-th row and $j$-th column, $\lceil\cdot\rceil$ denotes the ceiling operator, $\mathcal{B}/b$ denotes removing the element $b$ from the set $\mathcal{B}$, $\mathcal{O}(\cdot)$ denotes the big-O notation, $\max\{ \cdot\}$ and $\min\{\cdot\}$ denote the selection of the maximum and minimum values, respectively, from a given set, $\cup$ denotes the union of two sets, and $\arctan2(\cdot)$ is the two argument arctangent function.

\section{System Model}
In this section, we first present the 6DMA-BS model and its corresponding channel model with users in the uplink transmission. Then, we introduce a general user distribution model based on the NHPP.

\subsection{6DMA-BS Model}
\begin{figure}[t]
\centering
\setlength{\abovecaptionskip}{0.cm}
\includegraphics[width=3.55in]{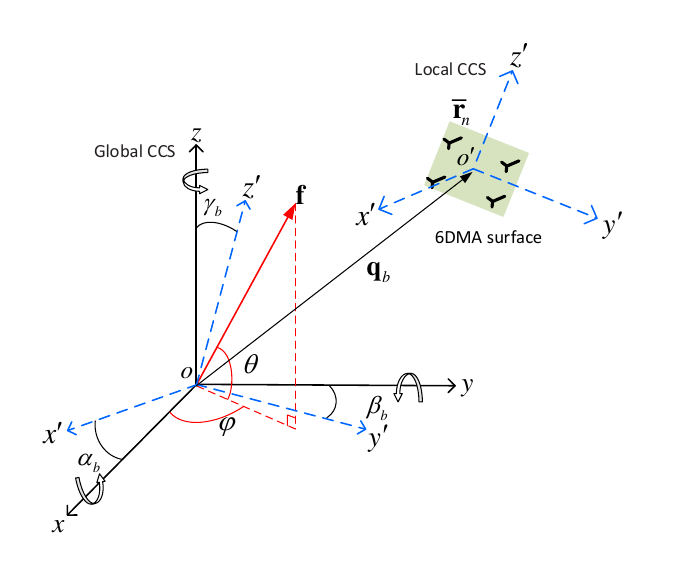}
\caption{Illustration of the geometry of the $b$-th 6DMA surface.}
\label{system}
\end{figure}
As shown in Fig. \ref{practical_scenario}, the proposed 6DMA-BS consists of $B$ 6DMA surfaces, denoted by the set $\mathcal{B} = \{1, 2, \ldots, B\}$. Each 6DMA surface is assumed to be a UPA with a given size, which consists of $N\geq 1$ antennas, denoted by the set $\mathcal{N} = \{1, 2, \ldots, N\}$. The 6DMA surfaces are connected to the CPU via extendable and rotatable rods embedded with flexible wires, and thus their 3D positions and 3D rotations can be adjusted by the CPU.
In particular, the position and rotation of the $b$-th 6DMA surface, $b\in\mathcal{B}$, can be characterized by six parameters, i.e., $\mathbf{q}_b$ for the 3D position and $\mathbf{u}_b$ for the 3D rotation (see Fig. \ref{system}), which are given by
\begin{align}\label{bb}
\mathbf{q}_b&=[x_b,y_b,z_b]^T\in\mathcal{C},\\
\mathbf{u}_b&=[\alpha_b,\beta_b,\gamma_b]^T,
\end{align}
where $\mathcal{C}$ denotes the given 3D space (e.g., a sphere or cube) at the BS in which the 6DMA surfaces can be flexibly positioned/rotated. We assume that $\mathcal{C}$ is a convex set which has a finite size. In the above, $x_b$, $y_b$ and $z_b$ represent the coordinates of the $b$-th 6DMA's center in the global Cartesian coordinate system (CCS) $o\text{-}xyz$, with the 6DMA-BS's reference position serving as the origin $o$; $\alpha_b\in[0,2\pi)$,  $\beta_b\in[0,2\pi)$ and $\gamma_b\in[0,2\pi)$ denote the rotation angles with respect to (w.r.t.) the $x$-axis, $y$-axis and $z$-axis, respectively.

Given $\mathbf{u}_b$, the following rotation matrix can be defined,
\begin{align}\label{R}
&\!\!\!\mathbf{R}(\mathbf{u}_b)=\mathbf{R}_{\alpha_b}\mathbf{R}_{\beta_b}
\mathbf{R}_{\gamma_b}\nonumber\\
&\!=\!\begin{bmatrix}
c_{\alpha_b}c_{\gamma_b} & c_{\alpha_b}s_{\gamma_b} & -s_{\alpha_b} \\
s_{\beta_b}s_{\alpha_b}c_{\gamma_b}-c_{\beta_b}s_{\gamma_b} & s_{\beta_b}s_{\alpha_b}s_{\gamma_b}+c_{\beta_b}c_{\gamma_b} & c_{\alpha_b}s_{\beta_b} \\
c_{\beta_b}s_{\alpha_b}c_{\gamma_b}+s_{\beta_b}s_{\gamma_b} & c_{\beta_b}s_{\alpha_b}s_{\gamma_b}-s_{\beta_b}c_{\gamma_b} &c_{\alpha_b}c_{\beta_b} \\
\end{bmatrix},\!\!
\end{align}
with
\begin{align}
&\mathbf{R}_{\alpha_b}=\begin{bmatrix}
1 & 0 & 0 \\
0 & c_{\alpha_b} & -s_{\alpha_b} \\
0 & s_{\alpha_b} & c_{\alpha_b} \\
\end{bmatrix},~\\
&\mathbf{R}_{\beta_b}=\begin{bmatrix}
c_{\beta_b} & 0 & s_{\beta_b} \\
0 & 1 & 0 \\
-s_{\beta_b} & 0 & c_{\beta_b} \\
\end{bmatrix},~\\
&\mathbf{R}_{\gamma_b}=\begin{bmatrix}
c_{\gamma_b} & -s_{\gamma_b} & 0 \\
s_{\gamma_b} & c_{\gamma_b} & 0 \\
0 & 0 & 1 \\
\end{bmatrix},
\end{align}
denoting the rotation matrices w.r.t. each of the  $x$-axis, $y$-axis and $z$-axis, respectively, where $c_{x}=\cos(x)$ and $s_{x}=\sin(x)$ are defined for notational simplicity \cite{rot3}.

As shown in Fig. \ref{system}, each 6DMA surface's local CCS is denoted by $o'\text{-}x'y'z'$, with the surface center serving as the origin $o'$. The $x'$-axis is oriented along the direction of the normal vector of the 6DMA surface.
Let $\bar{\mathbf{r}}_{n}$ denote the position of the $n$-th antenna of the 6DMA surface in its local CCS. Then, the position of the $n$-th antenna of the $b$-th 6DMA surface in the global CCS, denoted by $\mathbf{r}_{b,n}\in \mathbb{R}^3$, can be expressed as
\begin{align}\label{nwq}
\mathbf{r}_{b,n}(\mathbf{q}_b,\mathbf{u}_b)=\mathbf{q}_b+\mathbf{R}
(\mathbf{u}_b)\bar{\mathbf{r}}_{n},~n\in\mathcal{N},~b \in\mathcal{B}.
\end{align}

\begin{figure}[t!]
\centering
\setlength{\abovecaptionskip}{0.cm}
\includegraphics[width=2.1in]{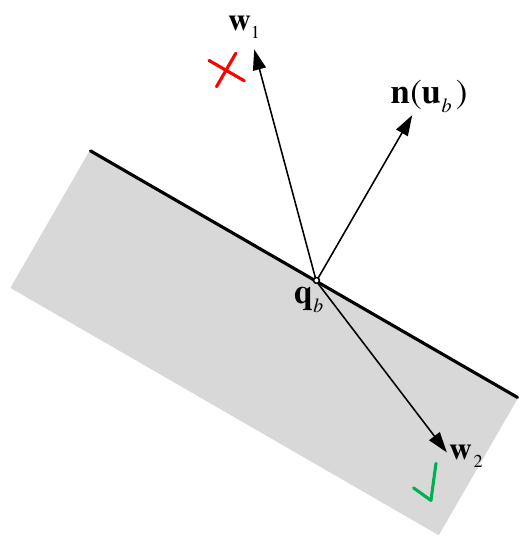}
\caption{The shaded set is the halfspace determined by $\mathbf{n}(\mathbf{u}_b)^T(\mathbf{w}-\mathbf{q}_b)\leq  0$. The vector $\mathbf{w}_1-\mathbf{q}_b$ makes an acute angle with $\mathbf{n}(\mathbf{u}_b)$, so $\mathbf{w}_1$ is not in the halfspace and does not avoid mutual signal reflection; whereas the vector $\mathbf{w}_2-\mathbf{q}_b$ makes an obtuse angle with $\mathbf{n}(\mathbf{u}_b)$, so $\mathbf{w}_2$ is in the halfspace and avoids mutual signal reflection.}
\label{halfspace}
\end{figure}

Next, we introduce three practical constraints on rotating/positioning 6DMA surfaces.
\subsubsection{Rotation Constraints to Avoid Signal Reflection}
6DMA surfaces must meet the following rotation constraints to avoid mutual signal reflections between any two 6DMA surfaces.
The outward normal vector of the $b$-th 6DMA surface can be defined as
\begin{align}
\mathbf{n}(\mathbf{u}_b)&=\mathbf{R}(\mathbf{u}_b)\bar{\mathbf{n}},\label{nuo}
\end{align}
where $\bar{\mathbf{n}}$ denotes the normal vector of the $b$-th 6DMA surface in the local CCS. For the $b$-th 6DMA surface, a hyperplane w.r.t. its center $\mathbf{q}_b$ and with its normal vector $\mathbf{n}(\mathbf{u}_b)$ is given by
\begin{align}\label{out}
\left\{\mathbf{w}|\mathbf{n}(\mathbf{u}_b)^T\mathbf{w}=c\right\},
\end{align}
where $c\in\mathbb{R}$. Note that the hyperplane in \eqref{out} can divide $\mathbb{R}^3$ into two halfspaces. One closed halfspace is defined as
\begin{align}\label{mut}
\left\{\mathbf{w}|\mathbf{n}(\mathbf{u}_b)^T(\mathbf{w}-\mathbf{q}_b)\leq  0\right\},
\end{align}
where $\mathbf{n}(\mathbf{u}_b)^T\mathbf{q}_b=c$. The above halfspace
consists of $\mathbf{q}_b$ as well as any vector that makes an obtuse angle
with the normal vector $\mathbf{n}(\mathbf{u}_b)$, as illustrated in
Fig. \ref{halfspace}. For the $b$-th 6DMA surface, when we set $\mathbf{w}$ in \eqref{mut} as the position of any antenna on a different
$j$-th 6DMA surface, i.e., $\mathbf{w}=\mathbf{r}_{j,n}, j \in \mathcal{B}/b, n\in\mathcal{N}$, the constraint to avoid mutual signal reflection between it and all the antennas of the $b$-th 6DMA surface is expressed as
\begin{align}\label{fcs}
\mathbf{n}(\mathbf{u}_b)^T(\mathbf{r}_{j,n}-\mathbf{q}_b)\leq 0,~\forall b,j \in \mathcal{B}, j\neq b, n\in\mathcal{N},
\end{align}
which ensures that none of the other 6DMA surfaces is positioned above the $b$-th 6DMA surface, thus preventing mutual signal reflections between any two 6DMA surfaces.

However, the above constraints need to be applied to all the antennas on all 6DMA surfaces. To reduce complexity, we relax the rotation constraint by setting $\mathbf{r}_{j,n}$ in \eqref{fcs} as the center position of the $j$-th 6DMA surface, i.e., $\mathbf{q}_{j}$. The rotation constraint in \eqref{fcs} then reduces to
\begin{align}\label{rcc}
\mathbf{n}(\mathbf{u}_b)^T(\mathbf{q}_{j}-\mathbf{q}_b)\leq  0,~\forall b ,j \in \mathcal{B}, j\neq b.
\end{align}

\subsubsection{Rotation Constraints to Avoid Signal Blockage}
\begin{figure}[t!]
\centering
\setlength{\abovecaptionskip}{0.cm}
\includegraphics[width=3.1in]{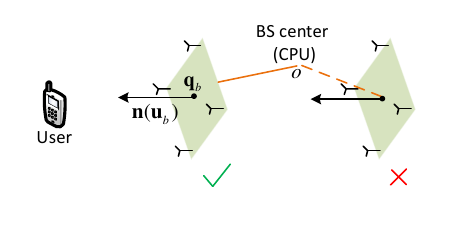}
\caption{Constraint on the rotation of each 6DMA surface
w.r.t. its attached rod.}
\label{rotationangle}
\end{figure}

To prevent each 6DMA surface from rotating towards the CPU of the BS which causes signal blockage, we impose a constraint on the rotation of each 6DMA surface w.r.t. its attached rod (see Fig. \ref{rotationangle}), which is given by
\begin{align}\label{dd}
\mathbf{n}(\mathbf{u}_b)^T\mathbf{q}_b\geq 0,~\forall b\in \mathcal{B}.
\end{align}

\subsubsection{Minimum-Distance Constraint}
We impose a minimum distance, denoted by $d_{\min}$,  between the centers of any pair of 6DMA surfaces to avoid their overlap as well as mutual coupling. This constraint is expressed as
\begin{align}\label{dss}
\|\mathbf{q}_b-
\mathbf{q}_{j}\|_2\geq d_{\min},~\forall b ,j \in \mathcal{B}, j\neq b.
\end{align}

\subsection{Channel Model}
For the 6DMA-BS, the channel between each 6DMA surface and a user in the network depends on not only the location of the user, but also the 3D position as well as 3D rotation of the 6DMA surface in general. In this subsection, we consider the uplink transmission and model the channels from each user (assumed to be equipped with a single FPA) to different antennas of all 6DMA surfaces.
\subsubsection{6D Steering Vector}
Let $\phi\in[-\pi,\pi]$ and $\theta\in[-\pi/2,\pi/2]$ denote the azimuth and elevation angles, respectively, of a signal arriving at the BS w.r.t. its reference position.
The pointing vector corresponding to the direction $(\theta, \phi)$ is thus defined as
\begin{align}\label{KM}
\mathbf{f}=[\cos(\theta)\cos(\phi), \cos(\theta)\sin(\phi), \sin(\theta)]^T.
\end{align}
To account for phase differences at different antennas on the same 6DMA surface,  by combining \eqref{nwq} and \eqref{KM}, the steering vector of the $b$-th 6DMA surface is a function of its position $\mathbf{q}_b$ and rotation $\mathbf{u}_b$, which is given by
\begin{align}\label{gen}
\mathbf{a}(\mathbf{q}_b,\mathbf{u}_b)=& \left[e^{-j\frac{2\pi}{\lambda}
\mathbf{f}^T\mathbf{r}_{b,1}(\mathbf{q}_b,\mathbf{u}_b)},
\cdots, e^{-j\frac{2\pi}{\lambda}\mathbf{f}^T
\mathbf{r}_{b,N}(\mathbf{q}_b,\mathbf{u}_b)}\right]^T,\nonumber\\
&~~~~~~~~~~~~~~~~~~~~~~~~~~~~~~~~~~~~~~~~~~b\in\mathcal{B},
\end{align}
where $\lambda$ denotes the carrier wavelength.

\subsubsection{Effective Antenna Gain}
Next, we derive the effective antenna gain for each 6DMA surface, which depends on the signal arriving angles $(\theta, \phi)$ as well as the rotation $\mathbf{u}_b$ of the 6DMA surface in general. In addition, it is heavily dependent on the radiation pattern of each antenna, which characterizes the antenna radiation power distribution over different directions\cite{3gpp}.
Although we assume that all the antennas of all 6DMA surfaces adopt the same antenna radiation pattern in this paper, each 6DMA surface has a different rotation in general, which results in different effective antenna gains over 6DMA surfaces. For convenience, we model the
antenna radiation pattern based on the local CCS of each 6DMA surface. In particular, we need to derive the elevation and azimuth angles of the signal direction w.r.t. the center of the $b$-th 6DMA surface in its local CCS, denoted by $(\tilde{\theta}_{b}, \tilde{\phi}_{b})$. To achieve this goal, we project $-\mathbf{f}$ in \eqref{KM} onto the $b$-th 6DMA surface in its local CCS to obtain\footnote{Note that in downlink communication, $-\mathbf{f}$ should be changed to $\mathbf{f}$.}
\begin{align}\label{df}
[\tilde{x}_{b},\tilde{y}_{b},\tilde{z}_{b}]^T&=-\mathbf{R}(\mathbf{u}_b)^{-1}
\mathbf{f}=-\mathbf{R}(\mathbf{u}_b)^T\mathbf{f}.
\end{align}
Based on \eqref{df}, $(\tilde{\theta}_{b}, \tilde{\phi}_{b})$ can be obtained as
\begin{align}
\tilde{\theta}_{b}&=\pi/2-\arccos(\tilde{z}_{b}),\label{pp}\\
\tilde{\phi}_{b}&=
\arccos\left(\frac{\tilde{x}_{b}}{\sqrt{\tilde{x}_{b}^2+\tilde{y}_{b}^2}}
\right)\times\eta(\tilde{y}_{b}) \label{cc}.
\end{align}
with
\begin{align}\label{ee}
\eta(\tilde{y}_{b})=\left\{\begin{matrix}
 1,~\tilde{y}_{b}\geq 0\\
 -1, ~\tilde{y}_{b}< 0
\end{matrix}\right.
\end{align}

Next, we define the effective gain of each antenna of the $b$-th 6DMA surface  in the scale of dBi in terms of the local-CCS signal angles $(\tilde{\theta}_{b}, \tilde{\phi}_{b})$ as $A(\tilde{\theta}_{b}, \tilde{\phi}_{b})$ (to be specified in Section V based on the practical antenna radiation pattern).
Then, the effective antenna gain for the $b$-th 6DMA surface in the linear scale  is defined as
\begin{align}\label{gm}
g(\mathbf{u}_b)=10^{\frac{A(\tilde{\theta}_{b}, \tilde{\phi}_{b})}{10}},
\end{align}
which is a function of the rotation $\mathbf{u}_b$ of the $b$-th 6DMA surface.

\subsubsection{Effective Channel}
In this paper, for the purpose of exposition and simplicity, we assume the line-of-sight (LoS) channel\footnote{
The channel model can be easily extended to a more general multipath channel with the user, where each signal path can be similarly modeled as the LoS path given in \eqref{uk}.} between any user's location and the 6DMA-BS, which can be expressed as
\begin{align}\label{uk}
\mathbf{h}(\mathbf{q},\mathbf{u})\!=\!\sqrt{\nu}\left[\!\sqrt{g(\mathbf{u}_1)}
\mathbf{a}(\mathbf{q}_1,\mathbf{u}_1)^T,
\!\cdots\!,\sqrt{g(\mathbf{u}_B)}\mathbf{a}(\mathbf{q}_B,\mathbf{u}_B)^T
\!\right]^T,
\end{align}
with
\begin{align}
{\mathbf{q}}&=[\mathbf{q}_1^T,\mathbf{q}_2^T,\cdots,\mathbf{q}_B
^T]^T\in \mathbb{R}^{3B\times 1},\label{lk}\\
\mathbf{u}&=[\mathbf{u}_1^T,\mathbf{u}_2^T,\cdots,\mathbf{u}_B^T]^T\in \mathbb{R}^{3B\times 1} \label{lk1}.
\end{align}
In the above, $\nu=\epsilon_0d^{-\varsigma
}$ is the path gain, where $\varsigma$ denotes the path loss exponent, $\epsilon_0$ represents the channel power at the reference distance $d_0=1$ meter (m), and $d>d_0$ denotes the distance between the user's location and the reference position of the 6DMA-BS.
\subsection{User Distribution}
As shown in Fig. \ref{practical_scenario}, we consider the uplink multiuser communications, where a random number of users, denoted by $K$, are  spatially distributed in a given cell and they transmit independent messages to the 6DMA-BS. We use the general NHPP to model the locations of the users in the cell \cite{flow2}. Without loss of generality, we consider a 3D cell coverage region $\mathcal{L}\in \mathbb{R}^{3}$ served by the 6DMA-BS. For any location  $\mathbf{z}\in\mathcal{L}$, we assume a given NHPP density function $\rho(\mathbf{z})$ (users/m$^3$). As a result, $K$ is a Poisson random variable with its mean given by
\begin{align}\label{muw}
\mu=\int_{\mathcal{L}}\rho(\mathbf{z})d\mathbf{z},
\end{align}
and probability mass function (PMF) given by
\begin{align}\label{pmf}
\mathrm{Pr}[K=\tilde{K}]=\frac{\mu^{\tilde{K}}}{\tilde{K}!}e^{-\mu},~ \tilde{K}=0,1,2, \cdots.
\end{align}

\section{Problem Formulation}
We consider the uplink transmission from $K$ single-antenna users to the 6DMA-BS, and the received signals at the BS are given by
\begin{align}\label{ly}
\mathbf{y}&=\mathbf{H}(\mathbf{q},\mathbf{u})
\mathbf{x}+\mathbf{n},
\end{align}
where $\mathbf{x}=\sqrt{p}[x_1, x_2,\cdots, x_{K}]^T\in \mathbb{C}^{K\times 1}$ with $x_k$ denoting the transmit signal of user $k$ with the average power normalized to one, and
$p$ representing the transmit power of each user (assumed to be identical for all users). $\mathbf{H}(\mathbf{q},\mathbf{u})=[\mathbf{h}_1(\mathbf{q},\mathbf{u}),\mathbf{h}_2(\mathbf{q},\mathbf{u}),\cdots,
\mathbf{h}_{K}(\mathbf{q},\mathbf{u})]\in \mathbb{C}^{NB\times K}$ denotes
the multiple-access channel from all $K$ users to all 6DMA surfaces at the BS with $\mathbf{h}_k(\mathbf{q},\mathbf{u})\in \mathbb{C}^{NB\times 1}$ denoting the channel from user $k, k\in\{1, 2, ..., K\}$, which is defined according to \eqref{uk} based on the pointing vector from user $k$, denoted by $\mathbf{f}_k$. $\mathbf{n}\sim\mathcal{CN}(\mathbf{0}_{NB},\sigma^2\mathbf{I}_{NB})$ denotes the complex additive white Gaussian noise (AWGN) vector at the BS with zero mean and average power $\sigma^2$.

Assuming perfect CSI at the BS, optimal Gaussian signaling and multiuser joint decoding, the achievable sum-rate of all users is given by \cite{david}
\begin{align}
C(\mathbf{q},\mathbf{u})&=\log_2 \det \left(\mathbf{I}_{NB}+\frac{1}{\sigma^2}\sum_{k=1}^{K}p\mathbf{h}_k
(\mathbf{q},\mathbf{u})
\mathbf{h}_k(\mathbf{q},\mathbf{u})^H\right)\nonumber\\
&=\log_2 \det \left(\mathbf{I}_{NB}+\frac{p}{\sigma^2}\mathbf{H}(\mathbf{q},\mathbf{u})
\mathbf{H}(\mathbf{q},\mathbf{u})^H\right), \label{aaa}
\end{align}
in bits per second per Hertz (bps/Hz). It is worth noting that different from the conventional multiuser channel with FPAs, the capacity of the 6DMA-enabled wireless channel given in \eqref{aaa} is
dependent on the 6D positions and rotations of all 6DMA surfaces, i.e., $\mathbf{q}$ and $\mathbf{u}$, which influence the effective channel
matrix $\mathbf{H}(\mathbf{q},\mathbf{u})$.

Note that the network capacity ${C}(\mathbf{q},
\mathbf{u})$ in \eqref{aaa} is a random variable due to the randomness in
the number of users $K$ and their random locations in the coverage region $\mathcal{L}$.
To characterize the average network capacity, we need to average out such randomness.
Thus, by applying the expectation w.r.t. the random channel $\mathbf{H}$ (due to the Poisson distributed $K$ and random user locations), we obtain the average network capacity as
\begin{align}
C_{\rm avg}=\mathbb{E}_{\mathbf{H}}\left[{C}(\mathbf{q},\mathbf{u})\right]. \label{pc0}
\end{align}
Since it is difficult to analytically derive the expectation in \eqref{pc0}, we apply the standard Monte Carlo method to obtain an approximation of $C_{\rm avg}$  \cite{monte3}. This involves generating $S$ independent realizations of the number of users, $K$, and their locations, and then averaging the corresponding achievable sum-rates over all realizations. Thus, the average network capacity in \eqref{pc0} can be approximated as
\begin{align}
\hat{C}(\mathbf{q},\mathbf{u})=\frac{1}{S}\sum_{s=1}^{S}
{C}_s(\mathbf{q},\mathbf{u}), \label{pc}
\end{align}
where ${C}_s(\mathbf{q},\mathbf{u})={C}(\mathbf{q},\mathbf{u})|
\mathbf{H}_s$ denotes the achievable sum-rate of the $s$-th realization given the corresponding user-BS channel $\mathbf{H}_s$.

Next, we aim to maximize the approximate network capacity of 6DMA-enabled wireless system by jointly optimizing the 3D positions $\mathbf{q}$ and 3D rotations $\mathbf{u}$ of all 6DMA surfaces at the BS, subject to their practical constraints given in Section II-A. Accordingly, the optimization problem is formulated as
\begin{subequations}
\label{MG3}
\begin{align}
\text{(P1)}~~&~\mathop{\max}\limits_{\mathbf{q},\mathbf{u}}~~
\hat{C}(\mathbf{q},
\mathbf{u})\\
\text {s.t.}~&~\mathbf{q}_i\in\mathcal{C}, ~\forall i \in \mathcal{B}, \label{M1}\\
~&~ \|\mathbf{q}_i-
\mathbf{q}_{j}\|_2\geq d_{\min},~\forall i ,j \in \mathcal{B}, ~j\neq i,\label{M2}\\
~&~ \mathbf{n}(\mathbf{u}_i)^T(\mathbf{q}_{j}-\mathbf{q}_i)\leq  0,~\forall i ,j \in \mathcal{B}, ~j\neq i, \label{M3}\\
~&~\mathbf{n}(\mathbf{u}_i)^T\mathbf{q}_i\geq 0, ~\forall i\in \mathcal{B}.\label{jM3}
\end{align}
\end{subequations}
where constraint \eqref{M1} guarantees that the center of each 6DMA surface is located in the given BS's 3D site space $\mathcal{C}$. As discussed in Section II-A, the minimum distance $d_{\min}$ in constraint \eqref{M2} avoids overlapping and coupling among 6DMA surfaces. Constraint \eqref{M3} avoids antenna mutual signal reflection, while constraint \eqref{jM3} prevents signal blockage by the CPU of the BS.

Note that problem (P1) is a non-convex optimization problem because the objective function is non-concave over the positions $\mathbf{q}$ and rotations $\mathbf{u}$ of 6DMA surfaces, as well as the constraints in \eqref{M2}, \eqref{M3} and \eqref{jM3} are non-convex. Moreover, the positions $\mathbf{q}$ are coupled with rotations $\mathbf{u}$ in the objective function of (P1), which makes their joint optimization a challenging task.

\section{Proposed Algorithm}
\subsection{Problem Decomposition and Alternating Optimization}
To solve (P1) efficiently, we first re-express its objective function in terms of the position $\mathbf{q}_b, b\in \mathcal{B}$ and rotation $\mathbf{u}_b, b\in \mathcal{B}$ of one 6DMA surface, with those of the other surfaces given. This will facilitate our proposed alternating optimization algorithm to solve (P1) subsequently.

Specifically, in the $s$-th Monte Carlo realization, we first rewrite the channel matrix $\mathbf{H}_s(\mathbf{q},\mathbf{u})\in \mathbb{C}^{NB\times K_s}$ as
\begin{align}\label{go}
\mathbf{H}_s(\mathbf{q},\mathbf{u})= \left[\mathbf{A}_s(\mathbf{q}_1,\mathbf{u}_1),
\mathbf{A}_s(\mathbf{q}_2,\mathbf{u}_2),
\cdots,\mathbf{A}_s(\mathbf{q}_B,\mathbf{u}_B)\right]^H \boldsymbol{\Sigma}_s,
\end{align}
with
\begin{align}
\boldsymbol{\Sigma}_s&=\mathrm{diag}\{\sqrt{\nu_1},\sqrt{\nu_2},\cdots,
\sqrt{\nu_{K_s}}\}\in \mathbb{C}^{K_s\times K_s},\!\!\!\!\!\\
\mathbf{A}_s(\mathbf{q}_b,\mathbf{u}_b)&=
\left[\sqrt{g_{1}(\mathbf{u}_b)}\mathbf{a}_{1}(\mathbf{q}_b,\mathbf{u}_b),\cdots,
\sqrt{g_{K_s}(\mathbf{u}_b)}\mathbf{a}_{K_s}(\mathbf{q}_b,\mathbf{u}_b)\right]^H\nonumber\\
&~~~~~~~~~~~~~~~~~~~~~~~~~~~\in \mathbb{C}^{K_s\times N},~b\in \mathcal{B},\label{qq}
\end{align}
where $K_s$ is the number of users in the $s$-th realization.

According to \eqref{uk}, $\nu_k$, $g_{k}(\mathbf{u}_b)$, and $\mathbf{a}_{k}(\mathbf{q}_b,\mathbf{u}_b)$ in the above equations respectively represent the channel path gain, effective antenna gain, and steering vector of the $b$-th 6DMA surface for the signal from the $k$-th user, $k=1,2,\cdots,K_s$.

Next, we define
\begin{align}\label{wr}
\mathbf{Q}_s(\mathbf{q},\mathbf{u})=\sqrt{p}\mathbf{H}_s(\mathbf{q},\mathbf{u})
\in \mathbb{C}^{NB\times K_s},
\end{align}
and denote the $b$-th sub-matrix of $\mathbf{Q}_s(\mathbf{q},\mathbf{u})^H$ by
$\mathbf{W}_s(\mathbf{q}_b, \mathbf{u}_b)$,
which is only determined by the position and rotation of the $b$-th 6DMA surface and can be expressed as
\begin{align}\label{pra}
\mathbf{W}_s(\mathbf{q}_b, \mathbf{u}_b)=\sqrt{p}\boldsymbol{\Sigma}_s^H
\mathbf{A}_s(\mathbf{q}_b,\mathbf{u}_b)\in \mathbb{C}^{K_s\times N}.
\end{align}
By combining \eqref{wr} and \eqref{pra}, the sum-rate given in \eqref{aaa} is rewritten as
\begin{align}
&\!\!C_s(\mathbf{q},\mathbf{u})=\log_2 \det \left(\mathbf{I}_{NB}+\frac{1}{\sigma^2}
\mathbf{Q}_s(\mathbf{q},\mathbf{u})\mathbf{Q}_s(\mathbf{q},\mathbf{u})^H\right)\nonumber\\
&\!\!\overset{(a)}{=}\log_2 \det \left(\mathbf{I}_{K_s}+\frac{1}{\sigma^2}
\mathbf{Q}_s(\mathbf{q},\mathbf{u})^H\mathbf{Q}_s(\mathbf{q},\mathbf{u})\right)\nonumber\\
&\!\!=\log_2 \det \left(\mathbf{I}_{K_s}+\frac{1}{\sigma^2}\sum_{b=1}^{B}
\mathbf{W}_s(\mathbf{q}_b,\mathbf{u}_b)
\mathbf{W}_s(\mathbf{q}_b,\mathbf{u}_b)^H\right),\!\!\! \label{wsx}
\end{align}
where the equality marked by $(a)$ holds due to  $\det(\mathbf{I}_{p}+\mathbf{B}\mathbf{C})=\det(\mathbf{I}_{q}+\mathbf{C}\mathbf{B})$ for $\mathbf{B}\in \mathbb{C}^{p\times q}$ and $\mathbf{C}\in \mathbb{C}^{q\times p}$. Note that the sum-rate given in \eqref{wsx} is expressed in terms of position and rotation variables of individual 6DMA surfaces, i.e., $\{\mathbf{q}_b,\mathbf{u}_b\}_{b=1}^B$.

Then, by removing $\mathbf{W}_s(\mathbf{q}_b, \mathbf{u}_b)$ from $\mathbf{Q}_s(\mathbf{q},\mathbf{u})^H$ in \eqref{wr}, the remaining $K_s \times N(B-1)$ sub-matrix can be denoted by
\begin{align}\label{QB}
\mathbf{Q}_{s,b}^H=&[\mathbf{W}_s(\mathbf{q}_1, \mathbf{u}_1),\cdots,\mathbf{W}_s(\mathbf{q}_{b-1}, \mathbf{u}_{b-1}),\mathbf{W}_s(\mathbf{q}_{b+1}, \mathbf{u}_{b+1}),\nonumber\\
&\cdots,\mathbf{W}_s(\mathbf{q}_B, \mathbf{u}_B)].
\end{align}
Given $\mathbf{Q}_{s,b}$, the sum-rate given in \eqref{wsx} can be rewritten as
\begin{align}\label{p2}
&\bar{C}_s(\mathbf{q}_b,\mathbf{u}_b)=\nonumber\\
&\log_2 \det \left(\mathbf{I}_{K_s}+\frac{1}{\sigma^2}\left(\mathbf{Q}_{s,b}^H\mathbf{Q}_{s,b}+
\mathbf{W}_s(\mathbf{q}_b,\mathbf{u}_b)
\mathbf{W}_s(\mathbf{q}_b,\mathbf{u}_b)^H\right)\right),
\end{align}
where $\mathbf{Q}_{s,b}^H\mathbf{Q}_{s,b}$ is a positive definite matrix regardless of $\mathbf{q}_b$ and $\mathbf{u}_b$. Thus, with given feasible $\mathbf{q}_j$ and $\mathbf{u}_j, j\in\mathcal{B}/b$ for the problem (P1), its objective function is simplified as a function of $\mathbf{q}_b$ and $\mathbf{u}_b$ only, i.e.,
\begin{align}\label{dc1}
\tilde{C}(\mathbf{q}_b,\mathbf{u}_b)=\frac{1}{S}\sum_{s=1}^{S}
\bar{C}_{s}(\mathbf{q}_b,\mathbf{u}_b),
\end{align}
which can be maximized subject to the constraints of (P1) pertaining to $\mathbf{q}_b$ and $\mathbf{u}_b$ in the order of $b=1,\cdots,B$ sequentially and then repeated in an iterative manner (i.e., alternating optimization). It is worth noting that at each of the above iterations, to reduce the computational complexity of matrix multiplication, the
matrix $\mathbf{Q}_{s,b}^H\mathbf{Q}_{s,b}$ in \eqref{p2} can be updated based on $\mathbf{Q}_{s,b-1}^H\mathbf{Q}_{s,b-1}, 2 \leq b \leq B$ by
\begin{align}\label{wm}
\mathbf{Q}_{s,b}^H\mathbf{Q}_{s,b}=\mathbf{Q}_{s,b-1}^H\mathbf{Q}_{s,b-1}
+\mathbf{M}_{A,b}\mathbf{M}_{B,b}^H,
\end{align}
where $\mathbf{M}_{A,b}=[\mathbf{W}_s(\mathbf{q}_{b-1}, \mathbf{u}_{b-1}), \mathbf{W}_s(\mathbf{q}_b, \mathbf{u}_b)]\in\mathbb{C}^{K_s\times 2N}$ and $\mathbf{M}_{B,b}=[\mathbf{W}_s(\mathbf{q}_{b-1}, \mathbf{u}_{b-1}), -\mathbf{W}_s(\mathbf{q}_b, \mathbf{u}_b)]\in\mathbb{C}^{K_s\times 2N}$.

\subsection{Optimization of $\mathbf{q}_b$}
In each iteration of the proposed alternating optimization, we first optimize $\mathbf{q}_b$ with given $\{\mathbf{q}_j\}_{j\in \mathcal{B}/b}$ and $\{\mathbf{u}_j\}_{j\in\mathcal{B}}$. From (P1), the resulted problem for optimizing $\mathbf{q}_b$ is expressed as
\begin{subequations}
\label{mm}
\begin{align}
\text{(P2-b)}~~&~\mathop{\max}\limits_{\mathbf{q}_b}
~\tilde{C}(\mathbf{q}_b,\mathbf{u}_b) \\
\text {s.t.}~&~ \mathbf{q}_b\in\mathcal{C}, \label{MM1}\\
~&~ \|\mathbf{q}_{b}-
\mathbf{q}_{j}\|_2\geq d_{\min},~\forall j \in \mathcal{B}/b,\label{MM2}\\
~&~ \mathbf{n}(\mathbf{u}_b)^T(\mathbf{q}_{j}-\mathbf{q}_b)\leq  0, ~j\in \mathcal{B}/b,\label{MM3}\\
~&~ \mathbf{n}(\mathbf{u}_j)^T(\mathbf{q}_{b}-\mathbf{q}_j)\leq  0,~ j\in \mathcal{B}/b,\label{MM33}\\
~&~\mathbf{n}(\mathbf{u}_b)^T\mathbf{q}_b\geq 0, \label{oMM33}
\end{align}
\end{subequations}
where \eqref{MM3} and \eqref{MM33} are derived from \eqref{M3} for a given $b$.
In the above, constraints \eqref{MM1}, \eqref{MM3}, \eqref{MM33}, and \eqref{oMM33} are convex, while the objective function of (P2-b) is non-concave and constraint \eqref{MM2} is non-convex over $\mathbf{q}_b$. Thus, it is difficult to obtain the globally optimal
solution for problem (P2-b) efficiently. In the following, we first transform
the non-convex constraint in \eqref{MM2} into a convex form, and then apply a feasible direction method, namely the conditional gradient method \cite{linear} to solve this problem.
\begin{figure}[t!]
\centering
\setlength{\abovecaptionskip}{0.cm}
\includegraphics[width=3.0in]{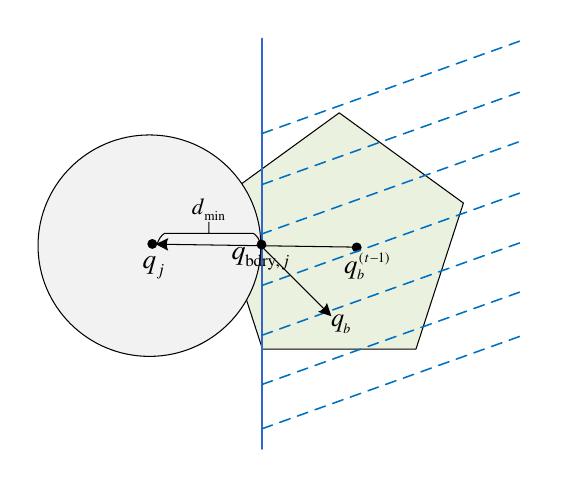}
\caption{Approximation of constraint \eqref{MM2}, where the dashed area is the halfspace determined by \eqref{bin1}.}
\label{feasible}
\end{figure}

As shown in Fig. \ref{feasible}, the feasible region of problem (P2-b) for $\mathbf{q}_b$ is a non-convex set in general (i.e., the green area). To transform the feasible region of (P2-b) into a convex set, we linearize the non-convex constraint in \eqref{MM2}. Specifically, in the $t$-th iteration, with the given $\mathbf{q}_b^{(t-1)}$ obtained in the last iteration and $\mathbf{q}_j, j\in\mathcal{B}/b$, we obtain the intersection of the vector $\mathbf{q}_b^{(t-1)} - \mathbf{q}_j$ with the spherical surface $\|\mathbf{q}_{b}-
\mathbf{q}_{j}\|_2=d_{\min}$ as the following boundary point (see Fig. \ref{feasible}),
\begin{align}\label{bin}
\!\!\mathbf{q}_{\mathrm{bdry},j}\!=\!\mathbf{q}_j-\frac{d_{\min}}{\|\mathbf{q}_j-\mathbf{q}_b^{(t-1)}\|_2}
\left(\mathbf{q}_j-\mathbf{q}_b^{(t-1)}\right), j\in \mathcal{B}/b.\!\!
\end{align}
Similar to the discussion in Section II-A, a hyperplane w.r.t. $\mathbf{q}_{\mathrm{bdry},j}$ and the vector $\mathbf{q}_b^{(t-1)} - \mathbf{q}_j$ is defined as
\begin{align}\label{outhy}
\left\{\mathbf{w}\big|(\mathbf{q}_j - \mathbf{q}_{\mathrm{bdry},j})^T\mathbf{w}=c\right\}.
\end{align}
Next, a closed halfspace divided by the above hyperplane is given by
\begin{align}\label{mut}
\left\{\mathbf{w}\big|\left(\mathbf{q}_j-\mathbf{q}_b^{(t-1)}\right)^T\left(\mathbf{w}
-\mathbf{q}_{\mathrm{bdry},j}\right)\leq 0\right\},
\end{align}
which is a halfspace consisting of $\mathbf{q}_{\mathrm{bdry},j}$ as well as any vector that makes an obtuse angle with the vector $(\mathbf{q}_j-\mathbf{q}_b^{(t-1)})$, as illustrated in Fig. \ref{feasible}.
By setting $\mathbf{w}=\mathbf{q}_b$ in \eqref{mut}, the non-convex constraint \eqref{MM2} can be approximated as the following linear inequality,
\begin{align}\label{bin1}
\left(\mathbf{q}_j-\mathbf{q}_b^{(t-1)}\right)^T\left(\mathbf{q}_b
-\mathbf{q}_{\mathrm{bdry},j}\right)\leq 0,~\forall j\in \mathcal{B}/b.
\end{align}

As a result, problem (P2-b) reduces to
\begin{subequations}
\label{rmm}
\begin{align}
\text{(P2-b-1)}~~&~\mathop{\max}\limits_{\mathbf{q}_b}
~\tilde{C}(\mathbf{q}_b,\mathbf{u}_b) \\
\text {s.t.}~&~ \eqref{MM1}, \eqref{MM3}, \eqref{MM33}, \eqref{oMM33}, \eqref{bin1}.
\end{align}
\end{subequations}

Since the feasible region of
problem (P2-b-1) is a convex set, it can be solved by using feasible direction methods. Typically, a feasible direction method starts with a feasible
vector $\mathbf{q}_b^{(0)}$ and generates a sequence of feasible vectors
$\{\mathbf{q}_b^{(t)}\}$ as
\begin{align}\label{gt}
\mathbf{q}_b^{(t)}=\mathbf{q}_b^{(t-1)}+\tau^{(t-1)}
(\tilde{\mathbf{q}}_b^{(t-1)}-\mathbf{q}_b^{(t-1)}),
\end{align}
where $\tau^{(t-1)}\in(0, 1]$ is the adaptive step size calculated by the Armijo rule \cite{arm}, $\tilde{\mathbf{q}}_b^{(t-1)}$ is a feasible vector different from $\mathbf{q}_b^{(t-1)}$, and $\tilde{\mathbf{q}}_b^{(t-1)}-\mathbf{q}_b^{(t-1)}$ is a feasible direction
(also known as a descent direction). Note that $\mathbf{q}_b^{(t)}$ in \eqref{gt} is always feasible
since the feasible region is a convex set. In the
following, we consider an efficient feasible direction method, i.e., the conditional gradient method \cite{linear}, to obtain $\tilde{\mathbf{q}}_b^{(t-1)}$ in \eqref{gt}.

In the conditional gradient method, the feasible vector $\tilde{\mathbf{q}}_b^{(t-1)}$ in \eqref{gt} is chosen as the
solution to the following optimization problem,
\begin{subequations}
\label{ormm}
\begin{align}
\text{(P2-b-2)}~~&~\mathop{\min}\limits_{\mathbf{q}_b}
~-\nabla_{\mathbf{q}_b}\tilde{C}(\mathbf{z}^{(t)})^T(\mathbf{z}-\mathbf{z}^{(t)}) \\
\text {s.t.}~&~ \eqref{MM1}, \eqref{MM3}, \eqref{MM33}, \eqref{oMM33}, \eqref{bin1},
\end{align}
\end{subequations}
where the gradient of function
$\tilde{C}(\mathbf{q}^{(t-1)}_b,\mathbf{u}_b))$ at the point  ${\mathbf{q}}^{(t-1)}_b$ is given by
\begin{align}\label{dff}
\!\!\!\!\![\nabla_{\mathbf{q}_b}\tilde{C}(\mathbf{q}^{(t-1)}_b, \mathbf{u}_b)]_j\!=\!&\lim_{\varepsilon\rightarrow 0}\frac{\tilde{C}(\mathbf{q}^{(t-1)}_b+\varepsilon\mathbf{e}^{j},
\mathbf{u}_b)\!-\!\tilde{C}(\mathbf{q}^{(t-1)}_b,\mathbf{u}_b)}{\varepsilon}, \nonumber\\
&~~~~~~~~~~~~~~~~~~~~~~~~~~~1 \leq j \leq 3,
\end{align}
where $\mathbf{e}^{j}\in \mathbb{R}^3$ is a vector with a one as the $j$-th element and zeros elsewhere. Note that problem (P2-b-2) is a linear optimization problem, which can be efficiently solved by using linprog \cite{lin}. The details of the conditional gradient algorithm for solving problem (P2-b) are presented in Algorithm 1.
\begin{algorithm}[t!]
\caption{Conditional Gradient Algorithm for Solving Problem (P2-b).}
\label{alg0}
\begin{algorithmic}[1]
\STATE \textbf{Input}: $B$, $N$, $\lambda$, maximum inner iteration number $T_{\mathrm{in}}$, step size $\tau_{\mathrm{ini}}$, and $\iota=10^{-2}$, $\delta=0.5$, $\{\mathbf{q}_j\}_{j\in \mathcal{B}/b}$, $\{\mathbf{u}_j\}_{j\in\mathcal{B}}$, and $\mathbf{u}_b^{(0)}$.  \\
\STATE {Initialization}: $t\leftarrow 0$.
 \WHILE{$t<T_{\mathrm{in}}$ }

    \STATE Compute the gradient value $\nabla_{\mathbf{q}_b}\tilde{C}(\mathbf{q}^{(t-1)}_b, \mathbf{u}_b)$ based on \eqref{dff} and set $\tau=\tau_{\mathrm{ini}}$;
     \STATE Obtain $\tilde{\mathbf{q}}_b^{(t-1)}$ by solving problem (P2-b-2);
     \STATE  Compute $\mathbf{q}_b^{(t)}=\mathbf{q}_b^{(t-1)}+\tau
(\tilde{\mathbf{q}}_b^{(t-1)}-\mathbf{q}_b^{(t-1)})$;
     \WHILE{$ \tilde{C}(\mathbf{q}^{(t)}_b, \mathbf{u}_b)-\tilde{C}(\mathbf{q}^{(t-1)}_b,\mathbf{u}_b)
       \!\!\!\!\!\!\!\!\!\!\!\!\!\!\!\!\!\!\!\!< \iota\tau\nabla_{\mathbf{q}_b}\tilde{C}(\mathbf{q}^{(t-1)}_b, \mathbf{u}_b)^T(\tilde{\mathbf{q}}_b^{(t-1)}-\mathbf{q}_b^{(t-1)})$}
     \STATE $\tau=\delta\tau$;
     \STATE  Compute $\mathbf{q}_b^{(t)}=\mathbf{q}_b^{(t-1)}+\tau
(\tilde{\mathbf{q}}_b^{(t-1)}-\mathbf{q}_b^{(t-1)})$;
     \ENDWHILE
     \STATE Update $t=t+1$;
     \ENDWHILE
 \STATE Return $\mathbf{q}_b^t$.
\end{algorithmic}
\end{algorithm}

\subsection{Optimization of $\mathbf{u}_b$}
Next, we optimize $\mathbf{u}_b$ with given $\{\mathbf{u}_j\}_{j\in \mathcal{B}/b}$ and $\{\mathbf{q}_j\}_{j\in\mathcal{B}}$ and the resulted problem for optimizing $\mathbf{u}_b$ is obtained from (P1) as
\begin{subequations}
\label{um}
\begin{align}
\text{(P3-b)}~~&~\mathop{\max}\limits_{\mathbf{u}_b}
~\tilde{C}(\mathbf{q}_b,\mathbf{u}_b),\\
\text {s.t.}~&~ \mathbf{n}(\mathbf{u}_b)^T(\mathbf{q}_j-\mathbf{q}_{b})\leq  0,~ j\in \mathcal{B}/b,\label{uuM3}\\
~&~\mathbf{n}(\mathbf{u}_b)^T\mathbf{q}_b\geq 0. \label{wgw}
\end{align}
\end{subequations}
The objective function of (P3-b)  is non-concave and constraints \eqref{uuM3} and \eqref{wgw} are non-convex, which makes it challenging to solve problem (P3-b) optimally.
Given the similarity in structure between problems (P2-b) and (P3-b),
we can apply the feasible direction method again to solve problem (P3-b), with the details given as follows.

First, we transform the non-convex constraints in \eqref{uuM3} and \eqref{wgw}  into convex forms. We denote $\mathbf{u}_b^{(t-1)}=[\beta_b^{t-1},\gamma_b^{t-1},\alpha_b^{t-1}]^T$ as the rotation solution after iteration $t-1$, and
\begin{align}\label{dg}
\Delta\mathbf{u}_b=\mathbf{u}_b-\mathbf{u}_b^{(t-1)}=[\Delta\beta_b, \Delta\gamma_b, \Delta\alpha_b]^T,
\end{align}
as the corresponding increments in the $t$-th iteration, where
$\Delta\beta_b=\beta_b-\beta_b^{t-1}$, $\Delta\gamma_b=\gamma_b-\gamma_b^{t-1}$, and $\Delta\alpha_b=\alpha_b-\alpha_b^{t-1}$. Then, the update for the rotation matrix at the current iteration can then be expressed as the product of the rotation matrix from the previous iteration, denoted as $\mathbf{R}(\mathbf{u}_b^{(t-1)})$, and the incremental rotation matrix, denoted as $\mathbf{R}(\Delta\mathbf{u}_b)$, that is \cite{sensor1},
\begin{align}\label{die}
\mathbf{R}(\mathbf{u}_b) = \mathbf{R}(\mathbf{u}_b^{(t-1)}) \mathbf{R}(\Delta\mathbf{u}_b).
\end{align}
Note that in \eqref{die}, as $\mathbf{R}(\mathbf{u}_b)$ is a unitary matrix belonging to the orthogonal group,
it is algebraically closed under the multiplication operation, but
not under addition \cite{sensor1}.

Since the angle changes $\Delta\mathbf{u}_b$ are very small in each iteration,
we can apply the following small-angle approximations \cite{rot3}:
\begin{align}
\cos( x)\rightarrow 1,\label{sub1}\\
\sin( x) \rightarrow  x,\label{sub2}
\end{align}
for $x\rightarrow0$.
By substituting $\mathbf{u}_b$ in \eqref{R} with $\Delta\mathbf{u}_b$ and then using the linearization approximations in \eqref{sub1} and \eqref{sub2}, we obtain the following linear approximation,
\begin{align}\label{ana}
\mathbf{R}(\Delta\mathbf{u}_b) \approx\begin{bmatrix}
1 & \Delta\gamma_b & -\Delta\alpha_b \\
-\Delta\gamma_b & 1 &\Delta\beta_b \\
\Delta\alpha_b &-\Delta\beta_b & 1 \\
\end{bmatrix}.
\end{align}

Substituting \eqref{die} and \eqref{ana} into $\mathbf{n}(\mathbf{u}_b)$ as defined in \eqref{nuo}, we can linearize the non-convex constraints \eqref{uuM3} and \eqref{wgw} as follows:
\begin{align}
&\bar{\mathbf{n}}^T\mathbf{R}(\Delta\mathbf{u}_b)^T\mathbf{R}(\mathbf{u}_b^{(t-1)})^T(\mathbf{q}_j-\mathbf{q}_{b})\leq  0,~ \forall j\in \mathcal{B}/b,\label{apcqq}\\
&\bar{\mathbf{n}}^T\mathbf{R}(\Delta\mathbf{u}_b)^T\mathbf{R}(\mathbf{u}_b^{(t-1)})^T\mathbf{q}_{b}\geq  0.\label{www3}
\end{align}
Consequently, problem (P3-b) reduces to
\begin{subequations}
\label{umqq}
\begin{align}
\text{(P3-b-1)}~~&~\mathop{\max}\limits_{\mathbf{u}_b}
~\tilde{C}(\mathbf{q}_b,\mathbf{u}_b),\\
\text {s.t.}~&~ \eqref{apcqq}, \eqref{www3}.
\end{align}
\end{subequations}
Now, the feasible region of problem (P3-b-1) is a convex set, making it solvable using feasible direction methods. Specifically, a feasible direction method for rotation optimization starts with a feasible vector $\mathbf{u}_b^{(0)}$ and generates a sequence of feasible vectors $\{\mathbf{u}_b^{(t)}\}$ as
\begin{align}\label{gtu0}
\mathbf{u}_b^{(t)}=\mathbf{u}_b^{(t-1)}+\tau^{(t-1)}
(\tilde{\mathbf{u}}_b^{(t-1)}-\mathbf{u}_b^{(t-1)}),
\end{align}
where $\tilde{\mathbf{u}}_b^{(t-1)}$ is a feasible vector, which can be chosen as the solution to the following optimization problem \cite{linear}
\begin{subequations}
\label{pj}
\begin{align}
\text{(P3-b-2)}~~&~\mathop{\min}\limits_{\mathbf{u}_b}
~-\nabla_{\mathbf{u}_b}\tilde{C}(\mathbf{q}_b,\mathbf{u}_b^{(t-1)})^T(\mathbf{u}_b
-\mathbf{u}_b^{(t-1)}) \\
\text {s.t.}~&~ \eqref{apcqq}, \eqref{www3},
\end{align}
\end{subequations}
where the gradient of function
$\tilde{C}(\mathbf{q}_b,\mathbf{u}^{(t-1)}_b)$ at the point ${\mathbf{u}}^{(t-1)}_b$ is given by
\begin{align}\label{dffr}
\!\!\!\!\!\!\![\nabla_{\mathbf{u}_b}\tilde{C}(\mathbf{q}_b,\mathbf{u}^{(t-1)}_b)]_j\!=\!&\lim_{\varepsilon\rightarrow 0}\frac{\tilde{C}(\mathbf{q}_b, \mathbf{u}^{(t-1)}_b+\varepsilon\mathbf{e}^{j})-\tilde{C}
(\mathbf{q}_b, \mathbf{u}^{(t-1)}_b)}{\varepsilon}, \nonumber\\
&~~~~~~~~~~~~~~~~~~~~~~~~~~~~~1 \leq j \leq 3.
\end{align}
Note that problem (P3-b-2) is a linear optimization problem, and can be efficiently solved by using linprog \cite{lin}. The details of the conditional gradient algorithm for solving problem (P3-b) are presented in Algorithm 2.
\begin{algorithm}[t!]
\caption{Conditional Gradient Algorithm for Solving Problem (P3-b).}
\label{alg1}
\begin{algorithmic}[1]
\STATE \textbf{Input}: $B$, $N$, $\lambda$, maximum inner iteration number $T_{\mathrm{in}}$, step size $\tau_{\mathrm{ini}}$, and $\iota=10^{-2}$, $\delta=0.5$, $\{\mathbf{u}_j\}_{j\in \mathcal{B}/b}$, $\{\mathbf{q}_j\}_{j\in\mathcal{B}}$, and $\mathbf{u}_b^{(0)}$.  \\
\STATE {Initialization}: $t\leftarrow 0$.
     \WHILE{$t<T_{\mathrm{in}}$ }
     \STATE Compute the gradient value $\nabla_{\mathbf{u}_b}\tilde{C}(\mathbf{q}_b,\mathbf{u}^{(t-1)}_b)$ based on \eqref{dffr} and set $\tau=\tau_{\mathrm{ini}}$;
     \STATE Obtain $\tilde{\mathbf{u}}_b^{(t-1)}$ by solving problem (P3-b-2);
    \STATE Compute $\mathbf{u}_b^{(t)}=\mathbf{u}_b^{(t-1)}+\tau
(\tilde{\mathbf{u}}_b^{(t-1)}-\mathbf{u}_b^{(t-1)})$;
         \WHILE{$ \tilde{C}(\mathbf{q}_b, \mathbf{u}_b^{(t)})-\tilde{C}(\mathbf{q}_b,\mathbf{u}_b^{(t-1)})
    \!\!\!\!\!\!\!\!\!\!\!\!\!\!\!\!\!\!\!\! <\iota\tau\nabla_{\mathbf{u}_b}\tilde{C}(\mathbf{q}_b,\mathbf{u}^{(t-1)}_b)^T(\tilde{\mathbf{u}}_b^{(t-1)}-\mathbf{u}_b^{(t-1)})$ or $\mathbf{u}_b^t$ is not feasible}
     \STATE $\tau=\delta\tau$;
     \STATE Compute $\mathbf{u}_b^{(t)}=\mathbf{u}_b^{(t-1)}+\tau
(\tilde{\mathbf{u}}_b^{(t-1)}-\mathbf{u}_b^{(t-1)})$;
     \ENDWHILE
     \STATE Update $t=t+1$;
       \ENDWHILE
 \STATE Return ${\mathbf{u}}_b^t$.
\end{algorithmic}
\end{algorithm}

\subsection{Initialization}
\begin{figure}[h!]
\centering
\setlength{\abovecaptionskip}{0.cm}
\includegraphics[width=2.6in]{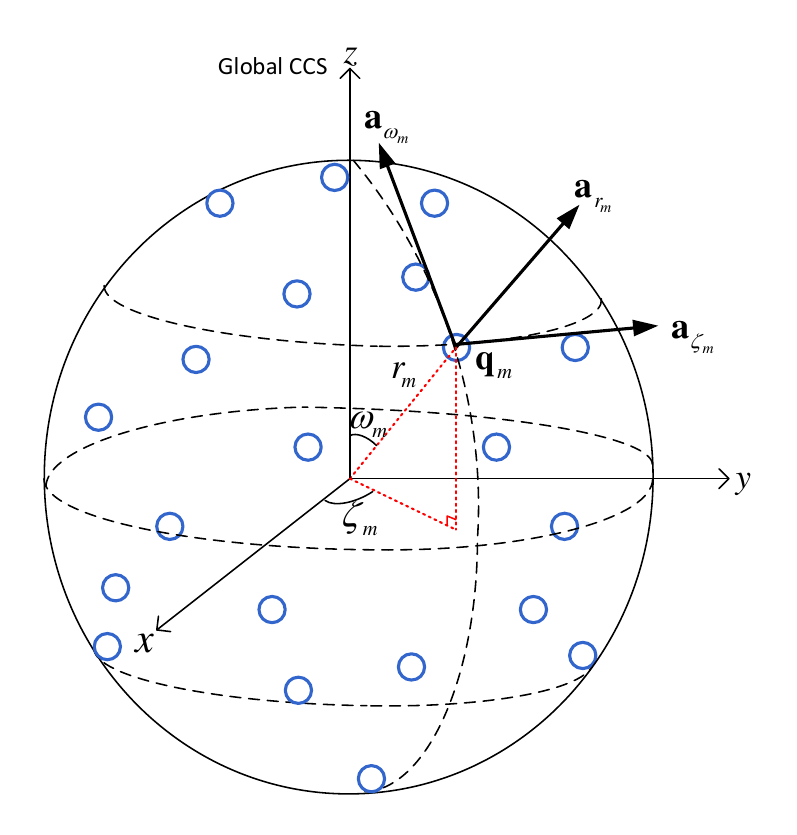}
\caption{Initial positions and rotations of the 6DMA surfaces (blue circles).}
\label{sphere}
\end{figure}
The above alternating optimization with the feasible direction method needs to select the initial values of $\mathbf{q}$ and $\mathbf{u}$ of 6DMA surfaces properly to ensure the good performance of the converged solution.
In this subsection, we propose a Fibonacci Sphere-based random initialization scheme. For selecting the initial positions within $\mathcal{C}$, it is desirable that they can well cover the BS coverage region $\mathcal{L}$.
In addition, the chosen positions and rotations must satisfy the practical rotation constraints in \eqref{M3} and \eqref{jM3} of (P1). Therefore, we propose to first uniformly generate $\varpi>B$ candidate positions on a spherical surface of the largest possible radius within the 6DMA-BS site space $\mathcal{C}$ by applying the Fibonacci Sphere scheme \cite{trip}, which can yield nearly uniform positions on the sphere that also satisfy the practical rotation constraints. Then, the initial positions are randomly selected from the candidate set $\Xi=\{1,2,\cdots, \varpi\}$.

Specifically, for each 6DMA surface's candidate position $\mathbf{q}_m, m\in \Xi$, we convert its Cartesian coordinates to the spherical coordinates $(r_m,\omega_m,\zeta_m)$ for convenience, where $r_m$, $\omega_m$, and $\zeta_m$ represent the radius, polar angle, and azimuthal angle, respectively, w.r.t. the reference position of the BS. For each location $\mathbf{q}_m$, there is a unique
rotation $\mathbf{u}_m$ obtained as follows. First, we orient the $x'$-axis along the direction of the radial basis vector $\mathbf{a}_{r_m}=[s_{\omega_m} c_{\zeta_m}, s_{\omega_m} s_{\zeta_m}, c_{\omega_m}]^T$ in the spherical coordinates. Similarly, we align the $y'$-axis along the direction of the azimuthal basis vector $\mathbf{a}_{\zeta_m}=[-s_{\omega_m} s_{\zeta_m}, s_{\omega_m} c_{\zeta_m}, 0]^T$, and the $z'$-axis along the direction of the polar basis vector $\mathbf{a}_{\omega_m}=[c_{\omega_m} c_{\zeta_m},c_{\omega_m} s_{\zeta_m},-s_{\omega_m}]^T$ (see Fig. \ref{sphere}).

Then, by substituting $\mathbf{q}_m$ and the corresponding basis vectors into \eqref{nwq}, the rotation matrix of the 6DMA surface located at the $m$-th candidate position can be determined as $\mathbf{R}(\mathbf{u}_m) = \left[\mathbf{a}_{r_m}, \mathbf{a}_{\zeta_m}, \mathbf{a}_{\omega_m}\right]$. Based on $\mathbf{R}(\mathbf{u}_m)$, rotation $\mathbf{u}_m$ corresponding to $\mathbf{q}_m$ can be obtained accordingly as \cite{rot3},
\begin{align}\label{hrhhre}
\mathbf{u}_m = \left[\begin{matrix}
\arctan2\left([\mathbf{R}(\mathbf{u}_m)]_{2,3},[\mathbf{R}(\mathbf{u}_m)]_{3,3}\right)\\ -\arcsin\left([\mathbf{R}(\mathbf{u}_m)]_{1,3}\right)\\ \arctan2\left([\mathbf{R}(\mathbf{u}_m)]_{1,2},[\mathbf{R}(\mathbf{u}_m)]_{1,1}\right)
\end{matrix}\right].
\end{align}
\begin{algorithm}[t!]
\caption{Alternating Optimization for Solving Problem (P1).}
\label{alg2}
\begin{algorithmic}[1]
\STATE \textbf{Input}: $B$, $N$, $\lambda$, maximum outer iteration number $T_{\mathrm{ou}}$.  \\
\STATE Initialize $\{\mathbf{q}_b\}_{b=1}^B$ and $\{\mathbf{u}_b\}_{b=1}^B$ using the Fibonacci Sphere-based random scheme.
\WHILE{$t<T_{\mathrm{ou}}$}
\FOR{$b = 1$ to $B$}
\STATE Obtain $\mathbf{Q}_{s,b}^H\mathbf{Q}_{s,b}$ via \eqref{wm};
    \STATE Given $\{\mathbf{q}_j\}_{j\in \mathcal{B}/b}$ and $\{\mathbf{u}_j\}_{j\in\mathcal{B}}$, solve problem (P2-b) by using Algorithm 1 to update $\mathbf{q}_b$;
       \ENDFOR
       \FOR{$b = 1$ to $B$}
      \STATE Obtain $\mathbf{Q}_{s,b}^H\mathbf{Q}_{s,b}$ via \eqref{wm};
      \STATE Given $\{\mathbf{u}_j\}_{j\in \mathcal{B}/b}$ and $\{\mathbf{q}_j\}_{j\in\mathcal{B}}$, solve problem (P3-b) by using Algorithm 2 to update $\mathbf{u}_b$;
     \ENDFOR
     \ENDWHILE
 \STATE Return $\mathbf{q}$ and $\mathbf{u}$.
\end{algorithmic}
\end{algorithm}

\subsection{Overall Algorithm}
With the solutions for problems (P2-b) and (P3-b) and the initialization scheme, the overall alternating optimization algorithm for solving problem (P1) is summarized in Algorithm 3. Specifically, from line 4 to line 7, we optimize the positions of all 6DMA surfaces, i.e., $\mathbf{q}$, by solving problem (P2-b) based on  Algorithm 1. Then, from line 8 to line 11, we optimize the rotations of all 6DMA surfaces, i.e., $\mathbf{u}$, by solving problem (P3-b) using Algorithm 2. The algorithm proceeds by iteratively solving the two subproblems until convergence.

Algorithm 3 is convergent since the alternating optimization and
gradient-based search ensure the objective value of (P1) to be non-decreasing over the iterations. Furthermore, since (P1) is a constrained problem, its objective value is upper-bounded by a finite value. In the following, we analyze the computational complexity of the proposed algorithm. Specifically, the computation of the gradients in \eqref{dff} of Algorithm 1 and \eqref{dffr} of Algorithm 2 each has a complexity of $\mathcal{O}(NB\bar{K}^2S)$ with $\bar{K} = \max(K_1, K_2, \ldots, K_S)$ due to the matrix multiplication. Thus, the overall complexity of the proposed alternating optimization algorithm is $\mathcal{O}(T_{\mathrm{ou}} T_{\mathrm{in}} NB\bar{K}^2S)$, where $T_{\mathrm{in}}$ and $T_{\mathrm{ou}}$ represent the maximum number of inner iterations in Algorithm 1 and 2, and outer iterations in Algorithm 3, respectively.

\section{Simulation Results}
\begin{figure}[t!]
\centering
\setlength{\abovecaptionskip}{0.cm}
\includegraphics[width=3.1in]{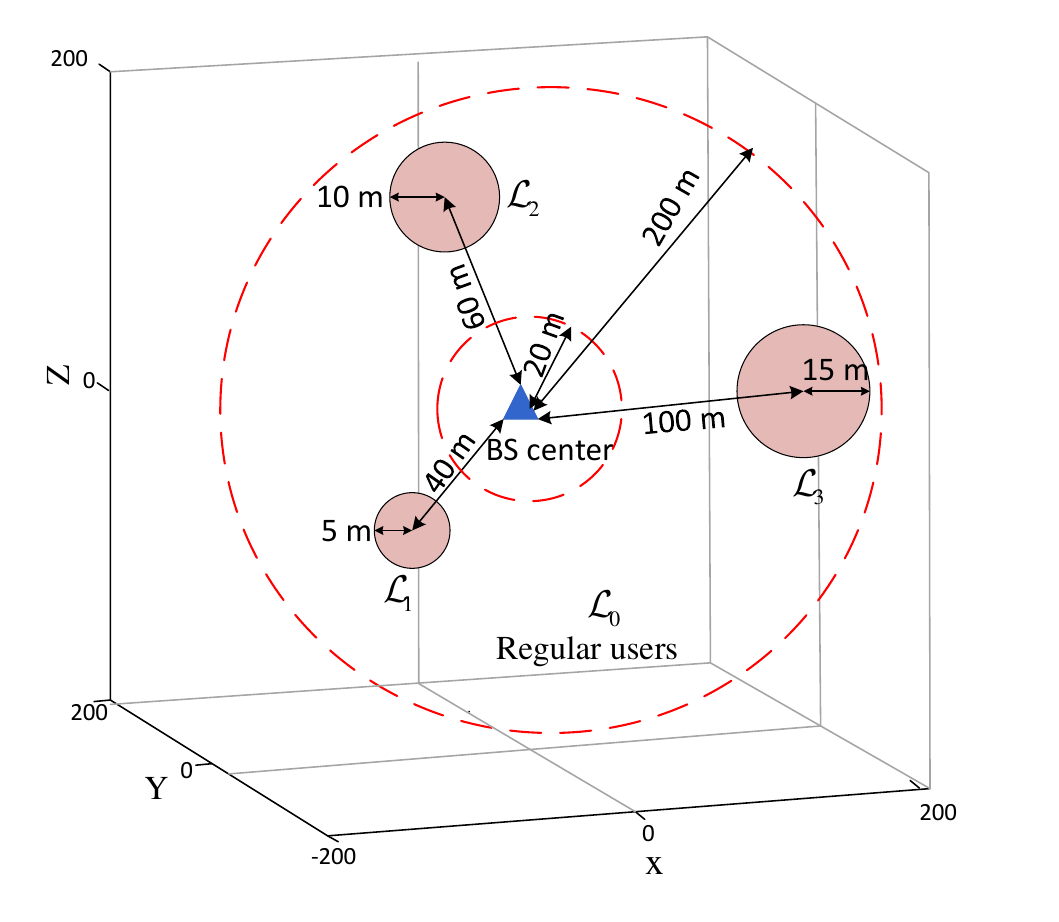}
\caption{Simulation setup for 6DMA system.}
\label{step}
\end{figure}
In this section, we provide numerical results to evaluate the performance of our proposed 6DMA-BS design and algorithm for maximizing the network capacity. As shown in Fig. \ref{step}, the users are located within a 3D coverage area $\mathcal{L}$. This area is defined as a 3D spherical annulus with radial distances ranging from 20 m to 200 m from the reference point of the 6DMA-BS. The coverage area $\mathcal{L}$ is divided into four subareas, which include $W=3$ non-overlapping hotspot areas, denoted by  $\mathcal{L}_w, w=1,2,3$, and the remaining area is denoted as $\mathcal{L}_0$, such that $\mathcal{L}=\mathcal{L}_0\cup(\cup_{w=1}^3\mathcal{L}_w)$.
The hotspot areas $\mathcal{L}_1$, $\mathcal{L}_2$, and $\mathcal{L}_3$ are defined as 3D spherical spaces centered at distances of 40 m, 60 m, and 100 m from the 6DMA-BS reference position, each with a radius of 5 m, 10 m, and 15 m, respectively.

The user density function is given by
\begin{align}\label{hq}
\rho(\mathbf{z})=\left\{\begin{matrix}
&\rho_0, &~\mathbf{z}\in\mathcal{L}_0,\\
&\rho_0+\rho_w, &~\mathbf{z}\in\mathcal{L}_w,  w\in\{1,2,3\},
\end{matrix}\right.
\end{align}
where $\rho_w\geq0$ and $\rho_0\geq0$ (in users/m$^3$) are constant user densities \cite{cluter}, which are set such that the average numbers of users in hotspot areas $\mathcal{L}_1$, $\mathcal{L}_2$, and $\mathcal{L}_3$ follow the ratio of  1:2:3. Furthermore, we define the {\emph{regular user ratio}} as $\xi=\frac{\int_{\mathcal{L}}\rho_0d\mathbf{z}}{\mu}$, with $\mu$ given in \eqref{muw}. $d_{\min}$ is set as $\frac{\sqrt{2}}{2}\lambda+\frac{\lambda}{2}$ with $\frac{\sqrt{2}}{2}\lambda$ being the diagonal length of UPA antenna surface. The main simulation parameters are provided in Table II, unless specified otherwise.
\begin{table}[!t]
\small
\caption{{Simulation Parameters.}}
\label{Table1}
\centering
\begin{tabular}{|c|c|c|c|c|}
\hline
Symbol & Description &Value \\
\hline
$N$&\makecell[c]{Number of antennas of \\each 6DMA surface}&4  \\
\hline
$B$&Number of 6DMA surfaces & 16 \\
\hline
$\mathcal{C}$& 6DMA-BS site space & Cube with 1 m sides\\
\hline
$\mu$ & Average number of users & 35\\
\hline
$d$&  User-BS distance & 20-200 m\\
\hline
$\sigma^2$& Average noise power & -50 dBm \\
\hline
$\lambda$ &Carrier wavelength & 0.125 m \\
\hline
$p$& Transmit power of user & 40 mW\\
\hline
$W$& Number of hotspots & 3\\
\hline
$S$& \makecell[c]{Number of Monte \\Carlo realizations}
 & 100\\
\hline
$\bar{d}$& \makecell[c]{Minimum antenna spacing\\
on each 6DMA surface}
 & $\lambda/2$\\
\hline
$\xi$& \makecell[c]{Regular user ratio}
 & $0.2$\\
\hline
$\varpi$& \makecell[c]{Number of candidate positions \\
by the Fibonacci Sphere \\scheme for initialization}
 & 64\\
\hline
\end{tabular}
\end{table}

\begin{figure}[t!]
\centering
\setlength{\abovecaptionskip}{0.cm}
\includegraphics[width=2.6in]{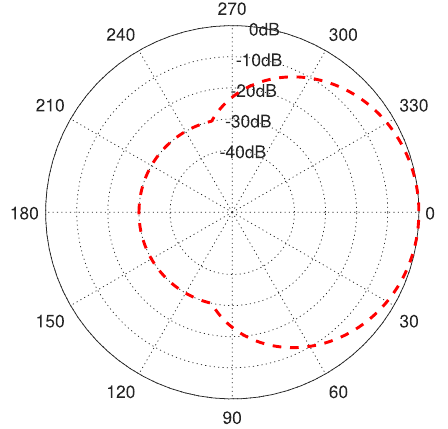}
\caption{Horizontal antenna radiation pattern.}
\label{pattern}
\end{figure}

The total number of users in each Monte Carlo realization is generated according to the PMF given in \eqref{pmf}. Note that the user locations within each subarea follow a uniform distribution. Hence, these locations can be easily generated using the random number generation method described in \cite{random}.
In addition, we model the effective antenna gain of the 6DMA-BS in \eqref{gm} according to the standard of 3GPP \cite{3gpp,radi3gpp}. Specifically, the antenna's radiation pattern includes both the horizontal and vertical patterns. According to \eqref{pp} and \eqref{cc}, the horizontal and vertical radiation patterns in dBi are respectively given by
\begin{align}
A_{\mathrm{H}}(\tilde{\phi}_{b})=-\min\left\{12\left(\frac{\tilde{\phi}_{b}}{\phi_{\mathrm{3dB}}}
\right)^2, G_s\right\},\label{AH}\\
A_{\mathrm{V}}(\tilde{\theta}_{b})=-\min\left\{12\left(\frac{\tilde{\theta}_{b}}{\theta_{\mathrm{3dB}}}
\right)^2,G_v\right\}, \label{Av}
\end{align}
where $\theta_{\mathrm{3dB}}$ and $\phi_{\mathrm{3dB}}$ both refer to the 3-dB beamwidth and take the same value of $65^\circ$, $G_s$ and $G_v$ are front-back ratio and sidelobe
level limit, respectively \cite{3gpp}.
Then, we can obtain the antenna gain $A(\tilde{\theta}_{b}, \tilde{\phi}_{b})$ in dBi for each pair of angles by combining the vertical and horizontal
radiation patterns as
\begin{align}\label{gm1}
A(\tilde{\theta}_{b}, \tilde{\phi}_{b})=G_{\max}-\min\left\{-[A_{\mathrm{H}}(\tilde{\phi}_{b})+A_{\mathrm{V}}(\tilde{\theta}_{b})],G_s\right\},
\end{align}
where $G_{\max}$ is the maximum directional gain of each
antenna element in the main lobe direction. In the simulation, $G_{\max}$ is set to 8 dBi, and both $G_s$ and $G_v$ are set to the identical value of 25 dBi.
In Fig. \ref{pattern}, we show the horizontal radiation pattern corresponding to the above model \cite{radi3gpp}.

For comparison, we consider the following benchmark schemes, all of which are based on a three-sector BS (i.e., a special case of the 6DMA with $B=3$ and $\lceil\frac{NB}{3}\rceil$ antennas on each surface). Each sector antenna covers approximately $120^{\circ}$. Moreover, the relative positions/rotations of the antennas on each sector antenna remain unchanged, while each sector antenna may independently change its center position or rotation.
\begin{itemize}
\item \textbf{FPA}: In this scheme, the 3D locations and 3D rotations of all sector antennas are fixed. The three sectors are all tilted towards the ground (with $\beta=15^{\circ}$ and $\alpha=0^{\circ}$).

\item \textbf{6DMA with circular movement}: In this scheme, the downtilts of three sector antennas are fixed (i.e., $\beta=15^{\circ}$ and $\alpha=0^{\circ}$). However, the center location of each sector antenna can move independently along a circular path that is parallel to the ground. We then apply the proposed algorithm to optimize the rotation of each sector antenna.

\item \textbf{6DMA with flexible-rotation only}: In this scheme, the center position of each sector antenna remains unchanged, and we apply the proposed algorithm to optimize the rotation of each sector antenna only.
\end{itemize}
\begin{figure}[t!]
\centering
\setlength{\abovecaptionskip}{0.cm}
\includegraphics[width=3.55in]{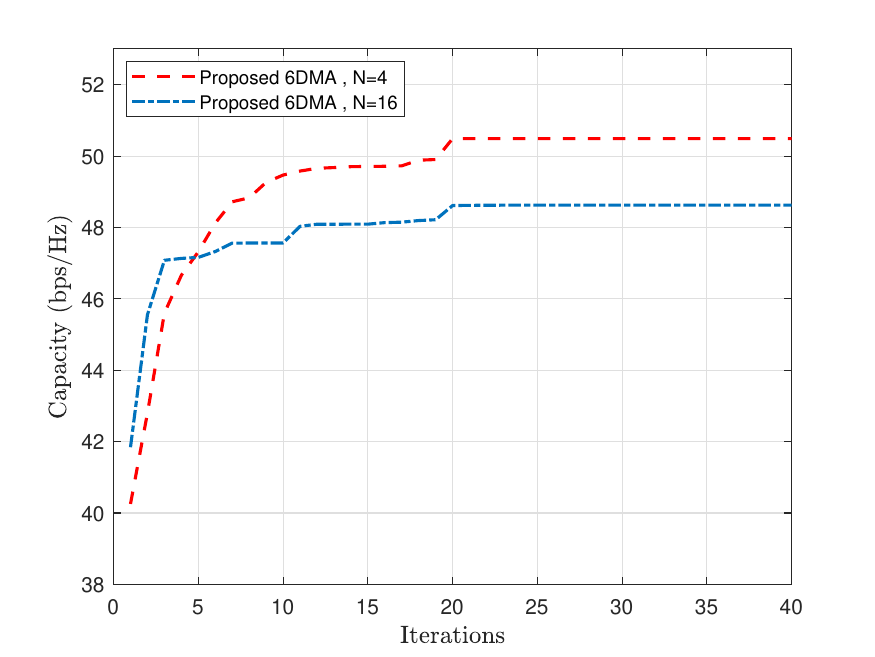}
\caption{Convergence behavior of the proposed algorithm.}
\label{iter}
\end{figure}

\begin{figure}[!t]
 \flushleft
\subfigure[$\xi=0$.]{
\begin{minipage}{8cm}
\centering
\includegraphics[scale=0.59]{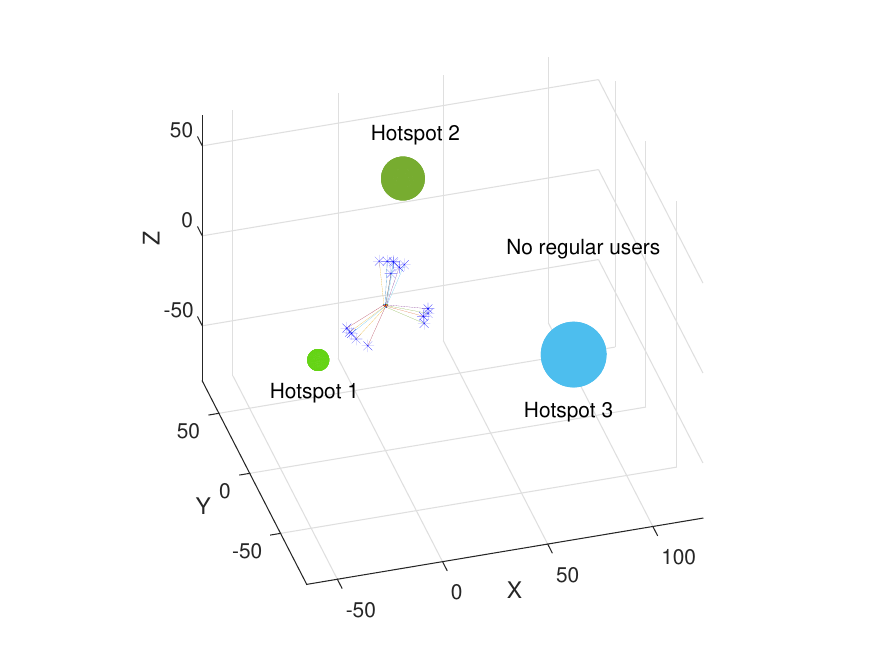}
\end{minipage}
}
\hspace*{-5mm}
\subfigure[$\xi=0.6$.]{
\begin{minipage}{5cm}
 \flushleft
\includegraphics[scale=0.41]{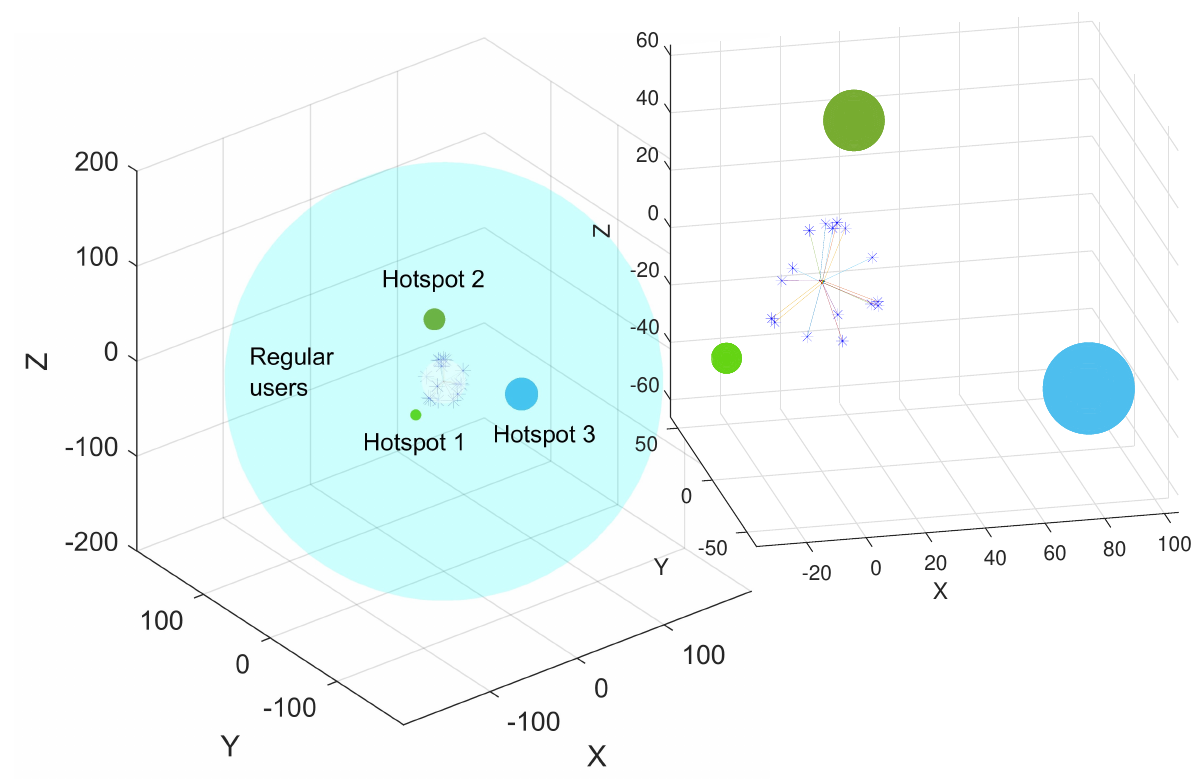}
\end{minipage}
}
\caption{The optimized positions/rotations of 6DMA surfaces using the proposed algorithm for different user distributions (the direction of each blue snowflake represents the normal vector of the corresponding 6DMA surface).}
\label{USERS}
\end{figure}

First, in Fig. \ref{iter}, with the total number of 6DMA-BS antennas fixed as 64, we illustrate the convergence behavior of the proposed algorithm (Algorithm 3)  under different numbers of antennas $N$ on each 6DMA surface. It is observed that the network capacity of the proposed algorithm with different $N$ increases over iterations and converges with fewer than 20 iterations. Notably, with $N=4$ (i.e., each 6DMA surface is a $2\times 2$ UPA), the performance surpasses that with $N=16$ (with each 6DMA surface as a $4\times$ 4 UPA). In other words, the network capacity of the proposed algorithm improves as the number of 6DMA surfaces increases due to more design flexibility. However, this gain comes at the cost of higher MA energy consumption and control complexity. Thus, a balanced trade-off between performance and cost needs to be considered for the 6DMA-BS configuration.

In Fig. \ref{USERS}, we illustrate the optimized positions/rotations of all 6DMA surfaces by the proposed algorithm based on the given user spatial  distribution. From Fig. \ref{USERS}(a), it is observed that in a network with only hotspot users (i.e., $\xi=0$), the 6DMA surfaces are generally oriented towards these hotspot areas. The allocation of 6DMA surfaces over different hotspots is mainly determined by both the BS-user distance and the number of users in each hotspot. In Fig. \ref{USERS}(b), we observe that in a network comprising both hotspot and regular users (with $\xi=0.6$), the 6DMA surfaces need to adjust their 6D positions and rotations to cater to both types of users.

\begin{figure}[t!]
\centering
\setlength{\abovecaptionskip}{0.cm}
\includegraphics[width=3.55in]{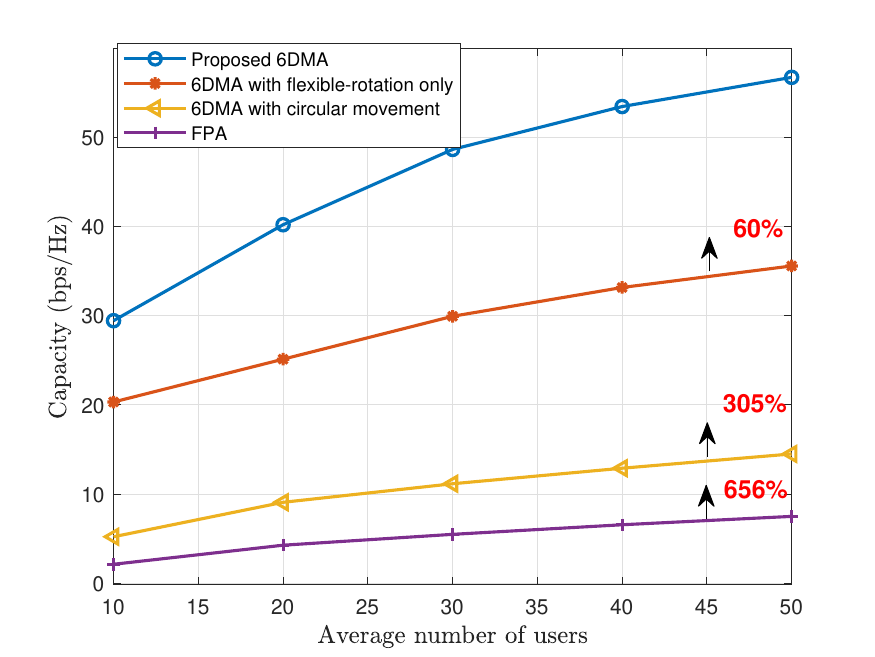}
\caption{Network capacity versus the average
number of users.}
\label{usernumber}
\end{figure}

Next, Fig. \ref{usernumber} shows the network capacity versus the average number of users, $\mu$. As expected, the network capacity with all considered schemes increases as
the number of users increases. The network capacity using the proposed scheme is substantially higher compared with the benchmark schemes. At $\mu=50$, the proposed scheme achieves 60\%, 305\%, and 656\% performance improvement over the 6DMA with flexible-rotation only, 6DMA with circular movement, and traditional FPA schemes, respectively. Such performance gains are attributed to the 6DMA-BS with the proposed algorithm by fully exploiting the flexible positions/rotations of 6DMA surfaces to maximize their array and spatial multiplexing gains based on the non-uniform user spatial distribution. In contrast, the FPA and MAs with limited/partial movability cannot take this advantage fully and their performance difference from
the proposed scheme becomes larger as the number of users increases. This indicates that the proposed BS design and algorithm are more appealing when the network is more heavily loaded.

\begin{figure}[t!]
\centering
\setlength{\abovecaptionskip}{0.cm}
\includegraphics[width=3.55in]{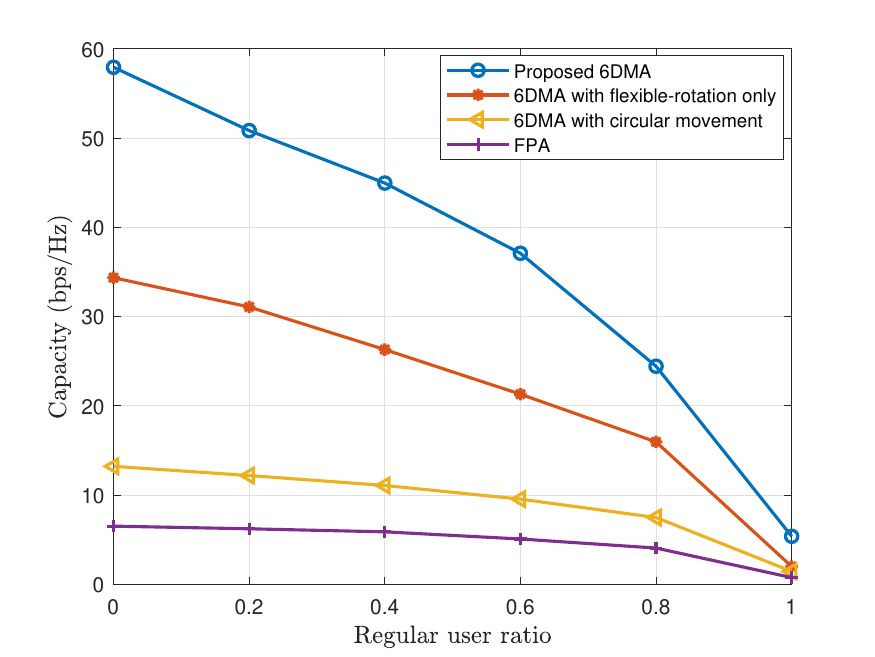}
\caption{Network capacity  versus the regular user ratio, $\xi$.}
\label{back}
\end{figure}

In Fig. \ref{back}, we evaluate the effect of the regular user ratio, $\xi$, on the network capacity under different schemes. We observe that the performance of the proposed algorithm, 6DMA with flexible-rotation only, and 6DMA with circular movement decreases as $\xi$ increases, i.e., the user spatial distribution approaches a uniform one. This is because when the users are more towards being  uniformly distributed, these schemes offer less advantages over the FPA scheme for serving the users isotropically. Furthermore, we observe that the proposed algorithm outperforms all the other schemes for all the values of $\xi$, while the performance gain decreases as $\xi$ increases towards one (i.e., with regular users only). The above results indicate that the proposed 6DMA-BS is more beneficial when the user spatial distribution exhibits more non-uniform and clustering (hot-spot) patterns.

\begin{figure}[t!]
\centering
\setlength{\abovecaptionskip}{0.cm}
\includegraphics[width=3.55in]{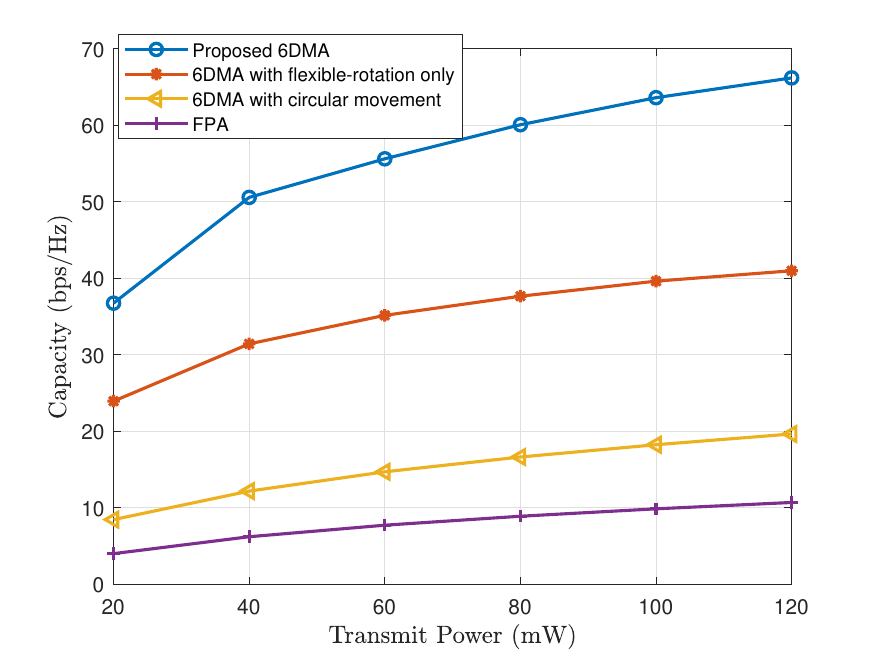}
\caption{Network capacity versus the
transmit power.}
\label{power}
\end{figure}

Finally, in Fig. \ref{power}, we show the network capacity of
the proposed and benchmark schemes versus user transmit power. It is observed that with the same transmit power, our proposed algorithm can achieve larger network capacity as compared to the benchmark schemes.
Furthermore, it is shown that the performance gap increases as the transmit power increases.
This is expected as the network capacity becomes more interference-limited as the transmit power or received signal-to-noise ratio (SNR) increases, and adjusting the positions/rotations of 6DMA surfaces based on the non-uniform user spatial distribution can effectively improve the multiuser-MIMO channel condition at the BS for more efficient interference mitigation.

\section{Conclusion}
In this paper, we proposed a novel 6DMA-enabled BS architecture for improving the wireless network capacity. Specifically, under the practical constraints on the 6DMA surfaces' movement, their 3D positions and 3D rotations were jointly optimized to maximize the network capacity based on the user spatial distribution. By employing the Monte Carlo simulation, alternating optimization, and conditional gradient methods, we proposed an efficient algorithm to solve this new problem with low complexity. Extensive simulation results under
practical setups were presented, which demonstrated that by employing the 6DMA-BS and proposed position/rotation optimization solution, the wireless network capacity  can be significantly improved, as compared to the benchmark BS architectures with traditional FPAs or MAs with limited/partial movability. Furthermore, it was shown that the capacity gains become more appealing when the user distribution is spatially non-uniform or the network traffic load/interference level is high. It is hoped that this paper will motivate a new approach for more efficiently deploying 6DMA in wireless networks by exploiting both their position and rotation DoFs for performance enhancement.


\begin{thebibliography}{1}
\bibitem{ruimimo}
A.~Goldsmith, S.~Jafar, N.~Jindal, and S.~Vishwanath, ``Capacity limits of
  {MIMO} channels,'' \emph{IEEE J. Sel. Areas Commun.}, vol.~21, no.~5, pp.
  684--702, Jun. 2003.


\bibitem{mimo3}
L.~Zheng and D.~Tse, ``Diversity and multiplexing: {A} fundamental tradeoff in
  multiple-antenna channels,'' \emph{IEEE Trans. Inf. Theory}, vol.~49, no.~5,
  pp. 1073--1096, May 2003.

\bibitem{SP8}
Q.~Spencer, A.~Swindlehurst, and M.~Haardt, ``Zero-forcing methods for downlink
  spatial multiplexing in multiuser {MIMO} channels,'' \emph{IEEE Trans. Signal
  Process.}, vol.~52, no.~2, pp. 461--471, Feb. 2004.

\bibitem{LA10}
E.~G. Larsson, O.~Edfors, F.~Tufvesson, and T.~L. Marzetta, ``Massive {MIMO}
  for next generation wireless systems,'' \emph{IEEE Commun. Mag.}, vol.~52,
  no.~2, pp. 186--195, Feb. 2014.

\bibitem{con9}
A.~Sayeed and N.~Behdad, ``Continuous aperture phased {MIMO}: Basic theory and
  applications,'' in \emph{Annu. Allerton Conf. Commun. Control. Comput.}, Sep.
  2010, pp. 1196--1203.

\bibitem{zenglens}
Y.~Zeng and R.~Zhang, ``Millimeter wave {MIMO} with lens antenna array: A new
  path division multiplexing paradigm,'' \emph{IEEE Trans. Commun.}, vol.~64,
  no.~4, pp. 1557--1571, Apr. 2016.

\bibitem{holo}
C.~Huang, S.~Hu, G.~C. Alexandropoulos, A.~Zappone, C.~Yuen, R.~Zhang, M.~D.
  Renzo, and M.~Debbah, ``Holographic {MIMO} surfaces for {6G} wireless networks:
  Opportunities, challenges, and trends,'' \emph{IEEE Wireless Commun.},
  vol.~27, no.~5, pp. 118--125, Oct. 2020.


\bibitem{qings}
Q.~Wu, S.~Zhang, B.~Zheng, C.~You, and R.~Zhang, ``Intelligent reflecting
  surface-aided wireless communications: A tutorial,'' \emph{IEEE Trans.
  Commun.}, vol.~69, no.~5, pp. 3313--3351, May 2021.

\bibitem{proc}
Q.~Wu \emph{et~al.}, ``Intelligent surfaces empowered wireless network: Recent
  advances and the road to {6G},'' \emph{arXiv preprint arXiv:2312.16918},
  2023.

\bibitem{shaos}
X.~Shao, C.~You, W.~Ma, X.~Chen, and R.~Zhang, ``Target sensing with
  intelligent reflecting surface: Architecture and performance,'' \emph{IEEE J.
  Sel. Areas Commun.}, vol.~40, no.~7, pp. 2070--2084, Jul. 2022.

\bibitem{nsr}
X.~Shao and R.~Zhang, ``Enhancing wireless sensing via a target-mounted
  intelligent reflecting surface,'' \emph{Nat. Sci. Rev.}, vol.~10, no.~8, p.
  nwad150, Jul. 2023.

\bibitem{shaotarget}
{X. Shao and R. Zhang}, ``Target-mounted intelligent reflecting surface for
  secure wireless sensing,'' \emph{IEEE Trans. Wireless Commun.}, early access,
  Feb. 2024.

\bibitem{net}
J.~Zhang, R.~Chen, J.~G. Andrews, A.~Ghosh, and R.~W. Heath, ``Networked {MIMO}
  with clustered linear precoding,'' \emph{IEEE Trans. Wireless Commun.},
  vol.~8, no.~4, pp. 1910--1921, Apr. 2009.

\bibitem{comp}
D.~Gesbert, S.~Hanly, H.~Huang, S.~Shamai~Shitz, O.~Simeone, and W.~Yu,
  ``Multi-cell {MIMO} cooperative networks: A new look at interference,''
  \emph{IEEE J. Sel. Areas Commun}, vol.~28, no.~9, pp. 1380--1408, Dec. 2010.

\bibitem{free}
H.~Q. Ngo, A.~Ashikhmin, H.~Yang, E.~G. Larsson, and T.~L. Marzetta,
  ``Cell-free massive {MIMO} versus small cells,'' \emph{IEEE Trans. Wireless
  Commun.}, vol.~16, no.~3, pp. 1834--1850, Mar. 2017.

\bibitem{m1}
R.~W. Heath, N.~Gonzalez-Prelcic, S.~Rangan, W.~Roh, and A.~M. Sayeed, ``An
  overview of signal processing techniques for millimeter wave {MIMO}
  systems,'' \emph{IEEE J. Sel. Topics Signal Process.}, vol.~10, no.~3, pp.
  436--453, Apr. 2016.

\bibitem{2m}
F.~Sohrabi and W.~Yu, ``Hybrid digital and analog beamforming design for
  large-scale antenna arrays,'' \emph{IEEE J. Sel. Topics Signal Process.},
  vol.~10, no.~3, pp. 501--513, Apr. 2016.

\bibitem{5m}
A.~F. Molisch, V.~V. Ratnam, S.~Han, Z.~Li, S.~L.~H. Nguyen, L.~Li, and
  K.~Haneda, ``Hybrid beamforming for massive {MIMO}: A survey,'' \emph{IEEE
  Commun. Mag.}, vol.~55, no.~9, pp. 134--141, Sep. 2017.


\bibitem{gold}
A.~Goldsmith, \emph{Wireless {Communications}}.\hskip 1em plus 0.5em minus
  0.4em\relax Cambridge university press, 2005.

\bibitem{david}
D.~Tse and P.~Viswanath, \emph{Fundamentals of {Wireless}
  {Communication}}.\hskip 1em plus 0.5em minus 0.4em\relax Cambridge university
  press, 2005.

\bibitem{til3}
C.~Weng, H.~Wang, K.~Li, and M.~N.~S. Swamy, ``Azimuth estimation for
  sectorized base station with improved soft-margin classification,''
  \emph{IEEE Access}, vol.~8, pp. 96\,649--96\,660, May 2020.

\bibitem{tilt}
N.~Dandanov, H.~Al-Shatri, A.~Klein, and V.~Poulkov, ``Dynamic
  self-optimization of the antenna tilt for best trade-off between coverage and
  capacity in mobile networks,'' \emph{Wireless Pers. Commun.}, vol.~92, pp.
  251--278, 2017.

\bibitem{9650760}
K.-K. Wong and K.-F. Tong, ``Fluid antenna multiple access,'' \emph{IEEE Trans.
  Wireless Commun.}, vol.~21, no.~7, pp. 4801--4815, Jul. 2022.

\bibitem{9264694}
W.~K. New, K.-K. Wong, H.~Xu, K.-F. Tong, C.-B. Chae, and Y.~Zhang, ``Fluid
  antenna system enhancing orthogonal and non-orthogonal multiple access,''
  \emph{IEEE Commun. Lett.}, vol.~28, no.~1, pp. 218--222, Jan. 2024.

\bibitem{wu}
L.~Zhu, W.~Ma, and R.~Zhang, ``Modeling and performance analysis for movable
  antenna enabled wireless communications,'' \emph{IEEE Trans. Wireless
  Commun.}, early access, Nov. 2023.

\bibitem{10243545}
W.~Ma, L.~Zhu, and R.~Zhang, ``{MIMO} capacity characterization for movable
  antenna systems,'' \emph{IEEE Trans. Wireless Commun.}, early access, Sept.
  2023.

\bibitem{yifei}
{Y. Wu, D. Xu, D. W. K. Ng, W. Gerstacker, and R. Schober}, ``Movable
  antenna-enhanced multiuser communication: Optimal discrete antenna
  positioning and beamforming,'' in \emph{IEEE Global Commun. Conf.
  (GLOBECOM)}, Dec. 2023, pp. 1--6.

\bibitem{qingmove}
G.~Hu, Q.~Wu, K.~Xu, J.~Si, and N.~Al-Dhahir, ``Secure wireless communication
  via movable-antenna array,'' \emph{IEEE Signal Process. Lett.}, vol.~31, pp.
  516--520, Jan. 2024.

\bibitem{review}
L.~Zhu and K.~K. Wong, ``Historical review of fluid antenna and movable
  antenna,'' \emph{arXiv preprint arXiv:2401.02362}, 2024.

\bibitem{3gpp}
3GPP, ``Technical specification group radio access network; study on {3D}
  channel model for {LTE},'' \emph{3rd Generation Partnership Project (3GPP),
  TR 36.873 V12.4.0}, 2017.

\bibitem{rot3}
J.~Diebel \emph{et~al.}, ``Representing attitude: {Euler} angles, unit
  quaternions, and rotation vectors,'' \emph{Matrix}, vol.~58, no. 15-16, pp.
  1--35, 2006.

\bibitem{flow2}
E.~Oh, K.~Son, and B.~Krishnamachari, ``Dynamic base station switching-on/off
  strategies for green cellular networks,'' \emph{IEEE Trans. Wireless
  Commun.}, vol.~12, no.~5, pp. 2126--2136, May 2013.

\bibitem{monte3}
I.~T. Dimov, \emph{Monte Carlo methods for applied scientists}.\hskip 1em plus
  0.5em minus 0.4em\relax World Scientific, 2008.

\bibitem{linear}
D.~P. Bertsekas, ``Nonlinear programming,'' \emph{J. Oper. Res. Soc.}, vol.~48,
  no.~3, pp. 334--334, 1997.

\bibitem{arm}
M.~Ahookhosh and S.~Ghaderi, ``On efficiency of nonmonotone {Armijo}-type line
  searches,'' \emph{Appl. Math. Model.}, vol.~43, pp. 170--190, 2017.

\bibitem{lin}
T.~Rocha, A.~Borges, S.~Paredes, and A.~Pinho, ``A {Matlab} tool for solving
  linear goal programming problems,'' in \emph{Experiment Int. Conf.}, Jun.
  2019, pp. 337--342.

\bibitem{sensor1}
J.~Jiang, G.~Wang, and K.~C. Ho, ``Sensor network-based rigid body localization
  via semi-definite relaxation using arrival time and doppler measurements,''
  \emph{IEEE Trans. Wireless Commun.}, vol.~18, no.~2, pp. 1011--1025, Feb.
  2019.

\bibitem{trip}
B.~Keinert, M.~Innmann, M.~S{\"a}nger, and M.~Stamminger, ``Spherical fibonacci
  mapping,'' \emph{ACM Trans. Graphics (TOG)}, vol.~34, no.~6, pp. 1--7, 2015.

\bibitem{cluter}
C.~Saha, H.~S. Dhillon, N.~Miyoshi, and J.~G. Andrews, ``Unified analysis of
  {HetNets} using {Poisson} cluster processes under max-power association,''
  \emph{IEEE Trans. Wireless Commun.}, vol.~18, no.~8, pp. 3797--3812, Aug.
  2019.

\bibitem{random}
R.~C. Cheng, ``Random variate generation,'' \emph{Handbook of Simulation}, pp.
  139--172, 1998.

\bibitem{radi3gpp}
M.~M.~U. Chowdhury, S.~J. Maeng, E.~Bulut, and S.~Guven, ``{3D} trajectory
  optimization in {UAV}-assisted cellular networks considering antenna
  radiation pattern and backhaul constraint,'' \emph{IEEE Trans. Aerosp.
  Electron. Syst.}, vol.~56, no.~5, pp. 3735--3750, Oct. 2020.

\end{thebibliography}
\end{document}